\documentclass[11pt]{article}
\pdfoutput=1
\usepackage{amsfonts,amsmath}
\usepackage{amssymb}
\usepackage{fancyhdr}
\usepackage{slashed}
\usepackage{graphicx}
\usepackage{subfigure}
\usepackage{color}

\usepackage{float} 

\usepackage{hyperref}
\usepackage[utf8]{inputenc}
\usepackage[titletoc]{appendix}

\usepackage{cleveref}

\def\hybrid{\topmargin -20pt    \oddsidemargin 0pt
        \headheight 0pt \headsep 0pt
        \textwidth 6.25in       
        \textheight 9.25in       
        \marginparwidth .875in
        \parskip 5pt plus 1pt   \jot = 1.5ex}

\hybrid

\def\baselinestretch{1.2}

\catcode`\@=11

\def\marginnote#1{}
\newcount\hour
\newcount\minute
\newtoks\amorpm
\hour=\time\divide\hour by60
\minute=\time{\multiply\hour by60 \global\advance\minute by-\hour}
\edef\standardtime{{\ifnum\hour<12 \global\amorpm={am}%
        \else\global\amorpm={pm}\advance\hour by-12 \fi
        \ifnum\hour=0 \hour=12 \fi
        \number\hour:\ifnum\minute<10 0\fi\number\minute\the\amorpm}}
\edef\militarytime{\number\hour:\ifnum\minute<10 0\fi\number\minute}

\def\draftlabel#1{{\@bsphack\if@filesw {\let\thepage\relax
   \xdef\@gtempa{\write\@auxout{\string
      \newlabel{#1}{{\@currentlabel}{\thepage}}}}}\@gtempa
   \if@nobreak \ifvmode\nobreak\fi\fi\fi\@esphack}
        \gdef\@eqnlabel{#1}}
\def\@eqnlabel{}
\def\@vacuum{}
\def\draftmarginnote#1{\marginpar{\raggedright\scriptsize\tt#1}}

\def\draft{\oddsidemargin -.5truein
        \def\@oddfoot{\sl preliminary draft \hfil
        \rm\thepage\hfil\sl\today\quad\militarytime}
        \let\@evenfoot\@oddfoot \overfullrule 3pt
        \let\label=\draftlabel
        \let\marginnote=\draftmarginnote
   \def\@eqnnum{(\theequation)\rlap{\kern\marginparsep\tt\@eqnlabel}%
\global\let\@eqnlabel\@vacuum}  }

\def\preprint{\twocolumn\sloppy\flushbottom\parindent 2em
        \leftmargini 2em\leftmarginv .5em\leftmarginvi .5em
        \oddsidemargin -.5in    \evensidemargin -.5in
        \columnsep .4in \footheight 0pt
        \textwidth 10.in        \topmargin  -.4in
        \headheight 12pt \topskip .4in
        \textheight 6.9in \footskip 0pt
        \def\@oddhead{\thepage\hfil\addtocounter{page}{1}\thepage}
        \let\@evenhead\@oddhead \def\@oddfoot{} \def\@evenfoot{} }

\def\numberbysection{\@addtoreset{equation}{section}
        \def\theequation{\thesection.\arabic{equation}}}

\def\underline#1{\relax\ifmmode\@@underline#1\else
        $\@@underline{\hbox{#1}}$\relax\fi}

\def\titlepage{\@restonecolfalse\if@twocolumn\@restonecoltrue\onecolumn
     \else \newpage \fi \thispagestyle{empty}\c@page\z@
        \def\thefootnote{\fnsymbol{footnote}} }

\def\endtitlepage{\if@restonecol\twocolumn \else \newpage \fi
        \def\thefootnote{\arabic{footnote}}
        \setcounter{footnote}{0}}  

\catcode`@=12
\relax

\def\figcap{\section*{Figure Captions\markboth
        {FIGURECAPTIONS}{FIGURECAPTIONS}}\list
        {Figure \arabic{enumi}:\hfill}{\settowidth\labelwidth{Figure
999:}
        \leftmargin\labelwidth
        \advance\leftmargin\labelsep\usecounter{enumi}}}
 \relax
\def\tablecap{\section*{Table Captions\markboth
        {TABLECAPTIONS}{TABLECAPTIONS}}\list
        {Table \arabic{enumi}:\hfill}{\settowidth\labelwidth{Table
999:}
        \leftmargin\labelwidth
        \advance\leftmargin\labelsep\usecounter{enumi}}}
 \relax
\def\reflist{\section*{References\markboth
        {REFLIST}{REFLIST}}\list
        {[\arabic{enumi}]\hfill}{\settowidth\labelwidth{[999]}
        \leftmargin\labelwidth
        \advance\leftmargin\labelsep\usecounter{enumi}}}
 \relax
\makeatletter
\newcounter{pubctr}
\def\publist{\@ifnextchar[{\@publist}{\@@publist}}
\def\@publist[#1]{\list
        {[\arabic{pubctr}]\hfill}{\settowidth\labelwidth{[999]}
        \leftmargin\labelwidth
        \advance\leftmargin\labelsep
        \@nmbrlisttrue\def\@listctr{pubctr}
        \setcounter{pubctr}{#1}\addtocounter{pubctr}{-1}}}
\def\@@publist{\list
        {[\arabic{pubctr}]\hfill}{\settowidth\labelwidth{[999]}
        \leftmargin\labelwidth
        \advance\leftmargin\labelsep
        \@nmbrlisttrue\def\@listctr{pubctr}}}
 \relax
\makeatother
\newskip\humongous \humongous=0pt plus 1000pt minus 1000pt

\newif\ifdtup

\relax

\def\be{\begin{equation}}
\def\ee{\end{equation}}
\def\ba{\begin{eqnarray}}
\def\ea{\end{eqnarray}}

\def\g{\gamma}

\def\d{\delta}

\def\e{\epsilon}

\def\p{\pi}

\def\m{\mu}

\def\n{\nu}

\def\l{\lambda}

\def\s{\sigma}
\def\S{\Sigma}

\def\cM{{\cal M}}

\def\cV{{\cal V}}

\def\cR{{\cal R}}

\def\cM{{\cal M}}  
  \def\cR{{\cal R}}
\def\cS{{\cal S}}  
\def\cV{{\cal V}}

\newcommand{\prt}[1]{{\left( {#1} \right)}}

\def\no{\noindent}

\def\IR{\relax{\rm I\kern-.18em R}}

\def\pp{\partial}

\newcommand{\ff}{\dfrac}

\def\IR{\relax{\rm I\kern-.18em R}}
\def\IL{\relax{\rm I\kern-.18em L}}

\def\inv{^{\raise.15ex\hbox{${\scriptscriptstyle -}$}\kern-.05em 1}}

\def\cM{{\cal M}}

\def\cV{{\cal V}}

\def\cR{{\cal R}}

\def\bea{\begin{eqnarray}}
\def\eea{\end{eqnarray}}
\newcommand{\eq}[1]{(\ref{#1})}
\def\nn{\nonumber}

\newcommand{\la}[1]{\label{#1}}

\def\g{\gamma}    
\def\d{\delta}    
\def\e{\epsilon}

\def\l{\lambda} 
\def\m{\mu} \def\n{\nu}

\def\p{\pi} 

\def\s{\sigma}  \def\S{\Sigma}

\definecolor{markcolor2}{rgb}{1,0,0}

\definecolor{markcolor3}{rgb}{0,1,0}

\newcommand{\Scal}{\mathcal{S}}

\begin{document}

\renewcommand{\theequation}{\thesection.\arabic{equation}}
\csname @addtoreset\endcsname{equation}{section}

\newcommand{\beq}{\begin{equation}}
\newcommand{\eeq}[1]{\label{#1}\end{equation}}
\newcommand{\ber}{\begin{eqnarray}}
\newcommand{\eer}[1]{\label{#1}\end{eqnarray}}
\newcommand{\eqn}[1]{(\ref{#1})}
\begin{titlepage}

\begin{center}

~
\vskip .7 cm

{\Large
\bf  Timelike Entanglement Entropy and Phase Transitions in non-Conformal Theories
  
}


\vskip 0.6in

 {\bf Mir Afrasiar,${}^{1,2}$
	Jaydeep Kumar Basak,${}^{1,2}$ and
	Dimitrios Giataganas${}^{1,2,3}$}
 \vskip 0.1in
 {\em
 	{\it ${}^1$
		Department of Physics, National Sun Yat-Sen University,
		Kaohsiung 80424, Taiwan\\}
	{\it ${}^2$
		Center for Theoretical and Computational Physics,
		Kaohsiung 80424, Taiwan\\}
	{\it ${}^3$
		Physics Division, National Center for
		Theoretical Sciences, Taipei 10617, Taiwan\\}
~ \\~\vskip .25in
 {\tt\href{mailto:mirhepth@gmail.com}{mirhepth@gmail.com}, \href{mailto:jkb.hep@gmail.com}{jkb.hep@gmail.com},\\
	\href{mailto:dimitrios.giataganas@mail.nsysu.edu.tw}
	{dimitrios.giataganas@mail.nsysu.edu.tw}
 }\\
 }

\vskip .1in
\end{center}

\vskip .2in

\centerline{\bf Abstract}
\noindent

We propose a holographic formalism for a timelike entanglement entropy in non-conformal theories. This pseudoentropy is a complex-valued measure of information, which, in holographic non-conformal theories, receives contributions from a set of spacelike surfaces and a finite timelike bulk surface with mirror symmetry.  We suggest a method of merging the surfaces so that the boundary length of the subregion is exclusively specified by holography. We show that in confining theories, the surfaces can be merged in the bulk at the infrared tip of the geometry and are homologous to the boundary region. The timelike entanglement entropy receives its imaginary and real contributions from the timelike and the spacelike surfaces, respectively. Additionally, we demonstrate that in confining theories, there exists a critical length within which a connected non-trivial surface can exist, and the imaginary part of the timelike entanglement entropy is non-zero. Therefore, the timelike entanglement entropy exhibits unique behavior in confining theories, making it a probe of confinement and phase transitions. Finally, we discuss the entanglement entropy in Euclidean spacetime in confining theories and the effect of a simple analytical continuation from a spacelike subsystem to a timelike one.

\no
\end{titlepage}
\vfill
\eject

\newpage
\tableofcontents

\newpage

\noindent


\def\baselinestretch{1.2}
\baselineskip 19 pt
\noindent


\setcounter{equation}{0}

\section{Introduction }
\label{sec_intro}

In recent years, there has been enormous progress in holographic entanglement measures. The starting point was the Ryu-Takayanagi prescription, where the entanglement entropy (EE) of a subregion of a boundary CFT is computed with a codimension-two extremal surface in the bulk \cite{Ryu:2006bv,Ryu:2006ef,Hubeny:2007xt}, with the endpoints of the bulk surface terminating at the boundary of the subregion. The subregion is assumed to factorize the total Hilbert space into $\mathcal{H}_{\text{tot}} = \mathcal{H}_A \otimes\mathcal{H}_B$, where $B$ is the complement of $A$. The entanglement entropy is naturally related to spacelike regions; in fact, the space coordinate can be thought of as emerging from the EE. A natural step forward is to extend this development by associating a new type of EE with the time coordinate. A way to define such an observable in quantum field theories, which we call timelike entanglement entropy (tEE),  is through an analytical continuation of the EE where the spacelike interval is analytically continued to a timelike region \cite{Doi:2023zaf}. In general, one expects a complex observable, with an imaginary contribution from the analytic continuation or, in holographic dual theory, from the analytically continued surfaces. The fact that the entropy becomes complex reflects the non-Hermiticity of the reduced density matrices involved. Therefore, the timelike entanglement entropy is not strictly a von Neumann entropy and can be considered more precisely as a pseudoentropy.

Holographically the tEE is expected to be related to the area of the timelike extremal surface, extending the correspondence between the spacelike intervals and the EE in an appropriate way. In static AdS${}_3$ spacetime, the standard EE slice can be Wick rotated into Euclidean AdS without any complications. This becomes evident in the dual two-dimensional quantum field theory, whereby Wick rotating the subregion $A$ associated with the Hilbert space $\mathcal{H}_A$, the tEE is defined by analytically continuing the EE to the timelike subsystem $A$. The analytic continuation on the length of $A$ introduces an imaginary unit $i$ in the logarithmic dependence of the standard EE, thus producing the imaginary part as
\be\label{res1}
S^{T}(A)=\frac{c}{3}\log \frac{T}{\e}+i\frac{c \pi}{6}~,
\ee
where $T$ is the length of the subsystem and is purely timelike. The imaginary part of the tEE in the two-dimensional theories is a characteristic property of the theory, i.e., it depends on its central charge and does not rely on the subregion length of $A$. This is further evidence that the tEE is a pseudoentropy. Despite this, the opportunity to obtain the central charge without the need for knowledge of the $T-$dependence of the observable via its imaginary part computation is itself fascinating. Moreover, we can argue that in two-dimensional conformal field theories, the constant nature of the imaginary part is universal, since any complexification in the logarithm will always produce a constant imaginary part. This expectation agrees with the findings in specific two-dimensional theories so far \cite{Doi:2023zaf,Chu:2023zah}.

A well defined notion of tEE should exist in time-dependent spacetimes and spacetimes with reduced symmetry. A certain special application is in dS/CFT \cite{Strominger:2001pn}, where the time coordinate emerges from a non-unitary Euclidean CFT. There, the entanglement entropy becomes complex-valued since there are no spacelike geodesics on the boundary to connect distinct points, and both spacelike and timelike surfaces contribute to the entropy. In fact, in this particular example, the timelike entanglement entropy has been linked to the pseudoentropy \cite{Doi:2022iyj}. Similarly, in Euclidean time-dependent asymptotically AdS spacetimes, the pseudoentropy is given by the regular minimal surface, generalizing the standard correspondence of the EE.

In generic quantum field theories, particularly those with less symmetry, the definition of the tEE from the field theory side with a Wick rotation is not as straightforward. For example, the tEE of \eqref{res1} can be computed alternatively by Wick rotating the coordinates of a Hamiltonian of a free scalar field theory with Lorentz invariance. In field theories where the analytic continuation exhibits complications for the EE this definition is more obscure. For instance in Lifshitz anisotropic field theories a holographic approach appears more feasible \cite{Basak:2023otu}. This is because the strict use of the Wick rotation can be avoided with a direct definition of the tEE in holographic dual theory and the complexification can be done following equivalent alternative ways. So far, all studies of the tEE have been concentrated in hyperbolic AdS, dS, in the special case of BTZ black holes, in BCFT and holographic Lifshitz theories \cite{Doi:2023zaf,Chu:2023zah,Basak:2023otu,Li:2022tsv,Narayan:2023zen,Guo:2024lrr}, and other related studies with the pseudoentropy \cite{Doi:2022iyj,Kanda:2023jyi,He:2024jog}.

In our current work, we propose the gravity dual definition of the tEE in non-conformal theories. The definition of the tEE in the holographic dual gravity side becomes tractable once a set of natural conditions on the surfaces are imposed. The contributions to the tEE are received from two different types of surfaces. The distinct spacelike surfaces between two points on the boundary, which are found by utilizing the usual parametrization that reduces the equations of motion to a one-parametric ordinary differential equation using translation invariance. The spacelike surface initiates from the boundary, where its exact boundary point is linked to the integration constants of the equations of motion. This surface does not have a turning point in the bulk, unlike the EE spacelike surfaces. In fact, we can think of them as being two separate pieces of spacelike surfaces that initiate from the boundary and each of them terminates on the IR wall of the bulk. The area of the spacelike surface is responsible for the real part of the timelike entanglement entropy. The timelike curve, which is a solution to the same equations of motion, is obtained by relaxing the condition that the surface ends at the boundary of the spacetime. We observe that, in practice, this corresponds to a complexification of the integration constant. The timelike surface remains in the bulk and has a mirror symmetry with a turning point towards the boundary of the space. The timelike part is responsible for the imaginary contribution to the tEE. We patch the timelike and spacelike surfaces by imposing certain natural conditions to obtain a smooth homologous surface to the boundary region.

Moreover, our proposal for the definition of the tEE in non-conformal theories takes into account the non-trivial dilaton field. In our prescription, we find all the extremal surfaces that contribute to the tEE by considering all possible solutions to the equations of motion, including the ones that are analytically continued.  In the initial proposal in conformal theories, the timelike and spacelike surfaces meet inside the bulk at infinite time direction and therefore the bulk data in general does not fix entirely the boundary endpoints of the surfaces unless additional information is used. The ambiguity directly reflects the uncertainty on the boundary subregion length $T$, since a constant shift is allowed via the integration constant. The ambiguity has been resolved for CFTs by the requirement that the surfaces are "merged" in such a way as to maintain the same boundary length after the analytic continuation,  i.e., the boundary length is determined in the EE, and the analytic continuation preserves it. Nevertheless, a natural expectation for the boundary length would be to be determined exclusively by the bulk construction when working on the holographic dual side. To this end, we identify a homologous, continuous smooth surface to the boundary subregion $A$ in such a way that fixes the boundary length uniquely. For confining theories, this surface is comprised of the spacelike surfaces that extend all the way from the boundary of the spacetime to the holographic IR wall, which in our spacetimes is the tip of the geometry; and a timelike surface that initiates from the IR wall of the geometry, extends towards the boundary direction, and returns to the IR wall, exhibiting mirror symmetry. The spacelike and timelike surfaces are merged to form a unique surface with no extra crossing points, homologous to a specific boundary length. We show that in confining theories with a compactified spatial direction of antiperiodic boundary conditions for fermions, this can be done smoothly at the tip of the geometry where the space shrinks.  As a side remark, we point out that our method does reproduce the AdS surfaces in conformal field theories that are known to contribute to the tEE, and the tEE in this case. In this sense, our holographic methods for the tEE apply effectively to both conformal and non-conformal theories, as expected

For the tEE in non-conformal systems, particularly when working in Lorentzian signatures, we observe a unique behavior associated with confinement. Therefore, we suggest that the tEE can serve as an order parameter for confinement/deconfinement phase transitions.  In confining theories, the tEE exhibits a maximum boundary length within which a connected non-trivial surface can exist, and its imaginary part of the tEE is non-zero. Beyond this critical length, the connected surface becomes trivial, taking the shape of the inverse $\Pi$, where the imaginary part lies entirely on the tip of the geometry and therefore contributes nothing. Both of these signals are unique to the tEE in confining theories. It's worth noting that for the standard EE, there has already been extensive discussion regarding its properties as a potential probe for phase transitions  \cite{Klebanov:2007ws,Kol:2014nqa,Jokela:2020wgs,Jeong:2022zea,Jokela:2023lvr,Baggioli:2023ynu,Fatemiabhari:2024aua}, as well as with he use of holographic c-functions defined via entanglement entropy \cite{Casini:2006es,Ryu:2006ef,Myers:2012ed,Chu:2019uoh,GonzalezLezcano:2022mcd} capable of identifying quantum phase transitions \cite{Baggioli:2020cld}. Our studies extend these developments to the tEE.

Moreover, when working in the Euclidean signature, it has been shown that only in certain conformal theories a simple analytic continuation of the subsystem length in the standard EE, can reproduce the tEE \cite{Doi:2023zaf}. In general, this is not expected to be true, since a naive analytic continuation from the spacelike system to a timelike does not automatically reflect in the desirable way on the conformal blocks. In confining theories, a naive analytic continuation of the EE exhibits qualitative similarities with the tEE results from the prescription described above. The EE undergoes a known phase transition, originating from the fact that the length of the boundary surface with respect to the turning point in the bulk is a double-valued function, which exhibits a maximal length \cite{Klebanov:2007ws}. For each turning point, there are two spacelike surfaces, one of which is favorable and stable. Before reaching the maximum boundary length for which a connected solution exists, a phase transition occurs on the minimal surfaces, where an inverse $\Pi$-shaped minimal surface becomes favorable. Therefore, following an alternative way by naively analytically continuing the EE, we observe a maximal length for which the connected solution exists, as well as a phase transition in the observable. It is interesting that the analytically continued EE computed in this way shares qualitative similarities with the tEE obtained by the computation of extremal surfaces in the Lorentzian signature. Nevertheless, the quantitative specifics of the two approaches differ.

The analytic computation of the tEE, like any other non-local observable in non-conformal theories, is generally possible under certain approximations. Therefore, in a part of our paper we resort to solving the spacelike and timelike surfaces numerically, and we describe our numerical methods in detail. Using these numerical solutions, we compute the tEE and extensively discuss our findings. Moreover, in a portion of our paper, we obtain analytic results via approximations using the binomial expansion \cite{Fischler:2012ca,Giataganas:2021jbj}. There are two limits where we can find analytical solutions: the limit of a small boundary length which corresponds to the timelike surface staying entirely around the tip of the IR geometry, and a limit where the turning point of the timelike surface reaches near the boundary of the spacetime. 
Both analytic and numerical methods provide valuable insights into the properties of the tEE in confining theories.

Our paper has the following structure. In Section 2, we propose the holographic definition of the tEE in non-conformal theories. In Section 3, we derive the equations of motion for the spacelike and timelike surfaces contributing to the tEE in a generic holographic spacetime. We determine the tEE via the extremal surfaces which this paper mostly focuses on, but we discuss also the tEE via a Wick rotation of EE. 
In Section 4, we apply our formalism to the solitonic background near-horizon limit of the D4-branes, which is known to exhibit confinement with a Wilson loop expectation value following an area law. Firstly, we discuss how we resolve the ambiguity of the boundary length in the holographic approach by defining the merging requirements of the homologous smooth surface. Then, we present analytical results in two limits: the limit near the tip of the geometry where the whole timelike surface lies closeby, and the limit near the boundary where the turning point of the timelike surface reaches the boundary of the geometry. In Section 5, we present numerical results for the tEE and discuss them. Finally, in Section 6, we recap and further discuss our findings. Our text is supported by Appendix A, where we compute the tEE in $AdS_3$ following our approach in comparison with the existing literature.

\section{Timelike Entanglement Entropy in non-Conformal Theories}\label{Euclid}

We consider in the string frame a generic metric with $d_0+1$ dimensions in the external space-time and $d_1$ dimension of the internal space-time $\cM$, with $d:=d_0+d_1$ and
\begin{equation}\label{genmet}
ds^2_{d+1}=g_{tt}(u)dt^{2}
	+g_{xx}(u)dx_i^2+g_{yy}(u)dy^{2}+g_{uu}(u)du^{2}~ + g_\cM(u) ds_{\cM}^2,
\end{equation}
where $u$ is the holographic direction, $t$ is the time direction, while we have defined the spatial $(d_0-2)$-directions $x_i$ and the $y$-direction separately as it will be convenient in the later stages of this article.  The boundary of the space is at $u\to \infty$. We allow the theory to be non-conformal with a generic non-constant dilaton $\phi(u)$. 

The entanglement entropy between a region $A$ of length $L$ and its complementary region is given holographically by the generalized Ryu-Takayangi formula \cite{Ryu:2006ef,Klebanov:2007ws}  
\be \la{def0}
S_A=\frac{1}{4 G_N^{(d+1)}}\int d\s^{d-1}  e^{-2 \phi} \sqrt{|g_{ind}^{(d-1)}|}~.
\ee
The induced metric can be taken as translational invariant along the directions in the internal and external spaces, and the resulting integral is effectively one-dimensional with $u$ dependence, obtained by minimizing the surfaces that approach the boundary of $A$.

The tEE for CFTs was recently introduced in \cite{Doi:2023zaf}. 
We propose the definition of the tEE for non-conformal theories as
\be \la{def1}
S_A^{(T)}=\ff{1}{4 G_N^{(d+1)}}\int d\s^{d-1} e^{-2 \phi} \sqrt{|g_{ind}^{(d-1),(T)}|}~,
\ee
where the induced string frame metric is the one appropriate for the timelike subsystem while in Lorentzian signature, and $G_N^{(d+1)}$ is the $(d+1)$-dimensional Newton constant. The formula  \eq{def1} takes into account the non-trivial contributions of the dilaton. The imaginary part of the action is generated by the timelike extremal surface that receives contributions from the internal spaces and the non-trivial dilaton. The spacelike surface contributes to the real part of the tEE.

In this work, using the definition \eq{def1} we study the tEE in confining theories and investigate its characteristics and its phase transitions.

\section{Formalism of the Timelike Entanglement Entropy}
\label{gen_timelike}

\subsection{Setup}

Let us consider in the metric \cref{genmet} a strip-like subsystem, $A= -T/2 < t < T/2$ of length $T$ in the time direction $t$ at the fixed $x_1=0$ slice on the asymptotic boundary. The timelike entanglement entropy can be computed by minimizing the area integral 
\begin{equation}\label{genSint}
	\mathcal{S}=
	\ff{\cV^{d_0-2}}{2G_N^{d+1}}	\int_{u_d}^{u_u} du~ V_{int}~g_{xx}^{\frac{d_0-3}{2}}~\sqrt{g_{yy}}~\sqrt{g_{uu}+g_{tt}\, t^\prime(u)^2}~,
\end{equation}
where $u_d$ and $u_u$ are the bottom and the upper bounds of the integral which we will specify later and depend on the type of spacetime and surface we consider. The upper bound is the boundary of the theory where the integral in \cref{genSint} contains the usual UV divergences, generated in the holographic side by the infinite proper length of the surfaces considered.  In \cref{genSint}, $\cV^{d_0-2}$ is the $(d_0-2)$ dimensional spatial volume, and $V_{int}$ is the volume of the internal manifold with the non-trivial contribution of the dilaton and metric:
\be\la{vaint}
V_{int}:=e^{-2 \phi}g_\cM(u){}^{\frac{d_1}{2}} \int_\cM \sqrt{|g_{int}|}:=e^{-2 \phi} g_\cM(u){}^{\frac{d_1}{2}}  \cV_\cM~,
\ee
where $\cV_\cM$ is the constant volume of the $\cM$. 
The equation of motion reads as
\begin{equation}\label{gentprime}
	t^\prime(u){}^2=\frac{c^2 ~g_{uu}}{ g_{tt}~(V_{int}^2~g_{yy}~g_{tt}~g_{xx}^{d_0-3}  -c^2)}~,
\end{equation}
where $c$ is an integration constant. Combining the \cref{genSint,gentprime} we can eliminate the derivatives and compute the on-shell action as,
\be 
\label{genSint2}
	\mathcal{S}=
	\ff{\cV^{d_0-2}}{2G_N^{d+1}}	\int_{u_d}^{u_u} du~ V_{int}^2~g_{xx}^{d_0-3}~g_{yy}~\sqrt{\ff{g_{tt}~g_{uu}}{V_{int}^2~g_{xx}^{d_0-3}~g_{yy}~g_{tt}-c^2}}~.
\ee 
Apart from the above mentioned surface connecting the endpoints of the subsystem, there is another solution of \cref{gentprime} which is described by two disconnected straight surfaces located at $t=\pm \ff{T}{2}$ and extended in all the spatial directions to the lowest point of the bulk which we call $u_k$. For these disconnected surfaces, the solution of the equations of motion reads $t^\prime(u)=0$ and the corresponding area integral is expressed as,
\begin{equation}\label{genSdis}
\mathcal{S}_{discon}= \ff{\cV^{d_0-2}}{2G_N^{d+1}} \int_{u_k}^{\infty} du~ V_{int}~g_{xx}^{\frac{d_0-3}{2}}~\sqrt{g_{yy}~g_{uu}}~.
\end{equation}

\subsection{Entanglement Entropy and Analytic Continuation}

Let us consider the metric in \cref{genmet} with Euclidean signature and the subsystem as $A= -T_E/2 < t_E < T_E/2$. For this case, we have a connected surface with mirror symmetry and a turning point we name $u_0$.  The entanglement entropy is equal to the area given by \cref{genSint}: $S_{con}=S$ with $u_d:=u_0$ and $u_u:=\infty$. The entanglement entropy is divergent and the UV divergence is the same as the divergence of the straight disconnected solutions \cref{genSdis}. Considering both the plausible solutions, we compute the difference between the area of these surfaces as,
\begin{align}
\mathcal{\hat S}&=\mathcal{S}_{con}-\mathcal{S}_{discon}\label{genSrenorm0}\\
	&=\frac{\cV^{d_0-2}}{2G_N^{d+1}} \Bigg[	\int_{u_0}^{\infty} du~ V_{int}\,g_{xx}^{\frac{d_0-3}{2}}\sqrt{g_{yy}~g_{uu}}\,\bigg( V_{int}\,g_{xx}^\frac{d_0-3}{2}\sqrt{g_{yy}}\,\sqrt{\frac{g_{tt}}{V_{int}^2\,g_{xx}^{d_0-3}\,g_{yy}\,g_{tt}-c^2}}-1\bigg)~\nn\\
	&\qquad\qquad\qquad\qquad\qquad\qquad\qquad\qquad \qquad\qquad\qquad\qquad\qquad~\,
    -\int_{u_k}^{u_0} du~ V_{int}\,g_{xx}^{\frac{d_0-3}{2}}\sqrt{g_{yy}\,g_{uu}}\,\Bigg]\label{genSrenorm},
\end{align}
where $\mathcal{\hat S}$ is free from any UV divergences. The sign of $\mathcal{\hat S}$ indicates the dominant option between $\mathcal{S}_{con}$ and $\mathcal{S}_{discon}$ as the surface with the smaller area is considered to be the natural Ryu-Takayanagi surface for any subsystem. 

The length of the subsystem can be computed by integrating \cref{gentprime},
\begin{align}\label{Tu0}
	T_E=2	\int_{u_0}^{\infty}	t_E^\prime(u)\, du~=2	\int_{u_0}^{\infty} \sqrt{\frac{c^2 \,g_{uu}}{ g_{tt}\left(V_{int}^2 \, g_{yy} \,g_{tt}\,g_{xx}^{d_0-3}  -c^2\right)}}\,du~.
\end{align}
The above function can be inverted to express the turning point $u_0$ as a function of $T_E$. Subsequently, replacing $u_0$ in \cref{genSrenorm}, $\mathcal{\hat S}$ can be written as a function of the length of the subsystem $A$. Finally, we perform a naive analytic continuation which replaces $T_E\to i\,T$ in the expression of $\mathcal{\hat S}$ to obtain the analytically continued EE.  Notice that it has been found that this type of analytic continuation is consistent with the tEE only for certain types of boundary subregions and CFTs \cite{Doi:2022iyj,Doi:2023zaf}. This is not surprising. In general, it is incorrect to simply analytically continue the EE expression aiming to obtain the tEE and a more careful treatment in most cases is required. In our work, we discuss this type of analytic continuation in the confining theories as well.

\subsection{Timelike Entanglement Entropy via Extremal Surface}\label{Lorentz}

Let us consider the metric \cref{genmet} with the Lorentzian signature and compute the tEE by analyzing the area of the corresponding extremal surfaces.
Considering a strip-like subsystem: $A= -T/2 < t < T/2$ of length $T$ in the time direction $t$ on $x_1=0$ slice at the asymptotic boundary $u=\infty$. The tEE $\mathcal{S}^T$ can be computed by \cref{def0} which gives the \cref{genSint} with  $\mathcal{S}^T=S$ and $u_d$ and $u_u$ depend on the type of the surface, spacelike versus timelike, which is determined by the signs of $c^2:=(i \tilde{c})^2$ and $\tilde{c}^2$, $\tilde{c}\in \cR$, as we describe below.

The denominator of the right-hand side of the differential equation that appears in \cref{gentprime} reads 
\be \la{denoa}
(-g_{tt})\,\left(\left(-g_{tt}\,\right)g_{xx}^{d_0-3}\,g_{yy}\,V_{int}^2 +c^2\right)~.
\ee
The derivative $t'(u)$ can be infinite in the bulk when $c^2<0$ and therefore  a turning point of the surface at $u_u=u_0$ exists given by
\be \la{gentur}
c^2=-(-g_{tt})\,g_{xx}^{d_0-3}\,g_{yy}\,V_{int}^2|_{u=u_0}~.
\ee
For later convenience, we redefine 
$c_0:=c/\cV_\cM$ to absorb the constant volume of the $\cM$ that appears in $V_{int}$ such that
\be \la{gentur2}
c_0^2=-(-g_{tt})\,g_{xx}^{d_0-3}\,g_{yy}\,\hat{V}^2_{int}|_{u=u_0}<0~.
\ee
Where $\hat{V}_{int}(u)=e^{-2 \phi} g_\cM(u){}^{\frac{d_1}{2}}$ is the $u-$dependent part of \eq{vaint} which includes the dilaton contribution and the $u-$dependent metric elements of the internal space, excluding the constant part volume of the internal space, e.g. if $\cM=S^4$ then  $V_{int}= \hat{V}_{int}(u) \cV_{S ^ 4} = \hat{V}_{int}(u)  \prt{8 \p^2/3}$. For this surface with $c_0$ given by \cref{gentur2}, we need to have $u<u_0$ in order to have a real solution. Therefore the solution $t(u)$ has a turning point at $u_u=u_0$ and extends from the turning point to the deep IR, which we denote as $u_d=u_k$, with a mirror symmetry. Consequently, this solution does not suffer from infinities since its proper length is finite. This specific surface obeys the boundary conditions $t^\prime(u_0)=\infty$ at the turning point $u_0$ and $t^\prime(u_k)=\pm\infty$ at the tip of the geometry $u=u_k$. This is the timelike surface we consider. 

Whereas, considering positive $c$, which we call it $\tilde{c}$ in the denominator of \cref{gentprime} we have
\be \la{deno}
(-g_{tt})\left(\left(-g_{tt}\right)\,g_{xx}^{d_0-3}\,g_{yy}\,V_{int}^2 +\tilde{c}^2\right)
\ee 
which is always positive for $\tilde{c}^2>0$ and the solution has no turning point.
Similar to \cref{gentur2} we define $\tilde{c}_0=\tilde{c}/\cV_{\cM}$ which can further be expressed as
\be \la{gentur3}
\tilde{c}_0^2=(-g_{tt})\,g_{xx}^{d_0-3}\,g_{yy}\,\hat{V}^2_{int}|_{u=u_0}>0~.
\ee
This is the spacelike surface.  The solution consists of a union of two different surfaces without any turning points, which extend from the boundary $u_u=\infty$ to the interior of the bulk $u_d=u_k$ and have infinite proper lengths, resulting in a UV divergence in the corresponding action. These two surfaces obey the boundary condition $t^\prime(\infty)=0$ at the boundary $u_u=\infty$ and $t^\prime(u_k)=\pm\infty$ at the interior $u=u_k$. 

Notice that both equations \eq{gentur2} and \eq{gentur3} are related by having the same $u_0$. Once the timelike surface is fixed, the spacelike surface is also fixed up to a constant which is specified by the additional information on how to take the union of the surfaces.

By computing the areas of the two surfaces, for $c_0^2<0$ and $\tilde{c}_0^2>0$, we define the timelike entanglement entropy to include all the possible contributions. The tEE is then given by
\begin{equation}\label{TEELor}
\mathcal{S}^T(u_0)=\mathcal{S}^T_{c_0}(u_0)+\mathcal{S}^T_{\tilde{c}_0}(u_0)~.
\end{equation}
The $\mathcal{S}^T_{\tilde{c}_0}$ contains UV divergences from the boundary $u_u=\infty$. To renormalize these divergences, we again consider $\mathcal{\hat S}^T$ by subtracting the area of the straight solution $\mathcal{S}^T_{discon}$ from \cref{TEELor}. Therefore, the observable related to the tEE, that is free of the UV divergences and considering all the possible solutions, can  be expressed as
\be\label{TEELor2}
	\mathcal{\hat S}^T(u_0)=\mathcal{S}^T_{c_0}(u_0)+\prt{\mathcal{S}^T_{\tilde{c}_0}(u_0)-\mathcal{S}^T_{discon}} ~.
\ee

We can obtain the subsystem lengths located at the asymptotic boundary $u_u=\infty$ for the two extremal surfaces corresponding to the solutions $c_0$ and $\tilde{c}_0$ discussed above. By integrating \cref{gentprime} for $c_0^2<0$ we obtain the subsystem $T_{c_0}$ for the extremal surface extending from the turning point $u_u=u_0$ to deep into the bulk $u_d=u_k$. Similarly, replacing $c$ by $\tilde{c}$ in \cref{gentprime} and integrating it with respect to the bulk direction $u$, we obtain the subsystem $T_{\tilde{c}_0}$ corresponding to the surface which extends from the asymptotic boundary $u_u=\infty$ to $u_d=u_k$. We can merge the surfaces in a smooth way so that the union is homologous to the boundary and has a well defined first derivative at the patching point. As we will demonstrate below at the IR wall of the confining geometry, which in our case is the tip of the geometry where the spatial compactified circle shrinks the timelike and spacelike surfaces momentarily become "null", and this is the only place we can smoothly merge them. We will find that for the confining theories under study, this requirement translates to the total subsystem length 
\be\label{T_tot_u0}
    T\left( u_0 \right) = T_{c_0} \left( u_0 \right) + T_{\tilde{c}_0}\left( u_0 \right)~.
\ee
Notice that for conformal theories, the same smoothness conditions for the surface lead to a modification of \cref{T_tot_u0} as it is explained in \cref{section:app1}.

In the following sections we will refer to $T_{c_0}$ as $T_{Im}$ and $T_{\tilde{c}_0}$ as $T_{Re}$ , with their respective significance discussed later. Furthermore, we express the turning point $u_0$ as a function of the total subsystem length $T$ by inverting \cref{T_tot_u0}. Replacing $u_0$ in the expression for $\mathcal{\hat S}^T$ in \cref{TEELor2}, we get the expression for UV free timelike entanglement entropy as a function of the subsystem length $T$. As a side remark, notice that some results would be qualitatively relevant for a boundary length \eq{T_tot_u0} defined with a subtraction instead of a summation, following an alternative (discarded in this paper) merging between the surfaces.

\section{Timelike Entanglement Entropy of a Confining Theory}\label{analytic_timelike}
In this section, we will explore a holographic background to compute the tEE.
Let us consider the near horizon limit of $N$ D4 branes, dual to the 5-d pure Yang-Mills theory. The reduction of the theory to four dimensions is achieved by compactifying one spatial direction $y$ on a circle $S^1$ with radius $R_0$ with anti-periodic boundary conditions for the fermions. The gravitational background in the string frame in the limit of small radius $R_0$ then reads
\begin{align}\label{Wmetric}
    ds^2=\left(\frac{u}{R}\right)^{3/2}\left(\eta_{\mu\nu}dx^\mu dx^\nu+f(u)dy^{2}\right)
	+\left(\frac{R}{u}\right)^{3/2}
	\left(\frac{du^2}{f(u)}+u^2 d\Omega_4^2\right)~,
\end{align}
where $\eta_{\mu\nu}dx^\mu dx^\nu=-dt^{2}+dx_i^{2}$ with $i=1,2,3$, denoting the uncompactified world volume along the D4-branes. In the above expression $R$ denotes the AdS length scale, $\Omega_4$ is the internal space of the unit four sphere, and the dilaton contribution and $f(u)$ are given by
\be \label{Wmetric2}
e^{\phi}= \prt{\frac{u}{R}}^{\frac{3}{4}}~, \qquad f(u) = 1-\left(\frac{u_k}{u}\right)^3~.
\ee
The background has a natural IR cut-off $u_k=4\pi\l/(9 R_0^2)$, which depends on the t'Hooft coupling $\l$ and the compactification radius $R_0$, and sets the non-singular tip of a cigar topology for the periodicity of $\d y:=2 \pi R_0$. This is the dual of the low temperature confining phase, referred as the D4 soliton, of an effectively four dimensional theory \cite{Witten:1998zw,Kruczenski:2003uq}.

\subsection{Extremal Surfaces for the Timelike Entanglement Entropy}

\begin{figure}[H]
	\centering
	\subfigure[ ]{\label{fig:sol1all}\includegraphics[width=71mm]{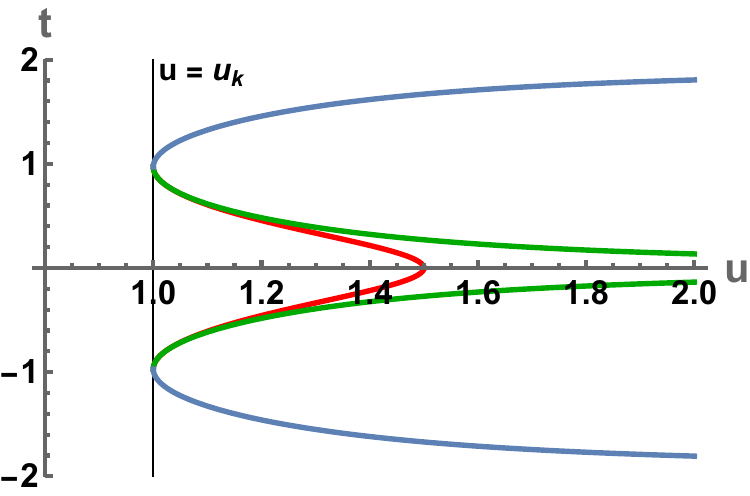}}\hspace{.5cm}
 	\subfigure[ ]{\label{fig:sol2all}  \includegraphics[width=71mm]{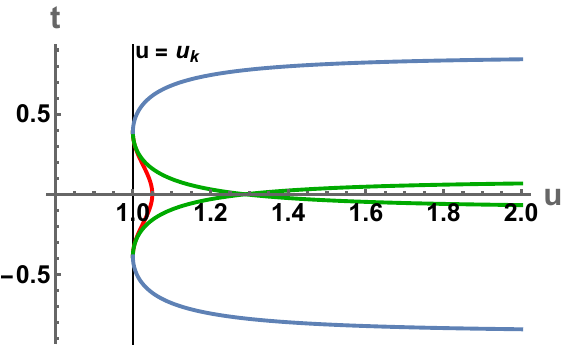}}
	\caption{The different ways to merge the timelike and spacelike surfaces obtained by solving \cref{eqtt1,eqtt2}. Following our prescription we choose the blue colored surfaces over the green ones,  since green surfaces violate the prescription. Moreover, the green surfaces intersect with each other when $u_0$ is near the tip of the geometry $u_k$. The blue and red surfaces are smoothly patched as depicted also in \cref{surface2}.}
	\label{surface}
\end{figure}

\begin{figure}[H]
	\centering
	\subfigure[ ]{\label{fig:sol1}\includegraphics[width=71mm]{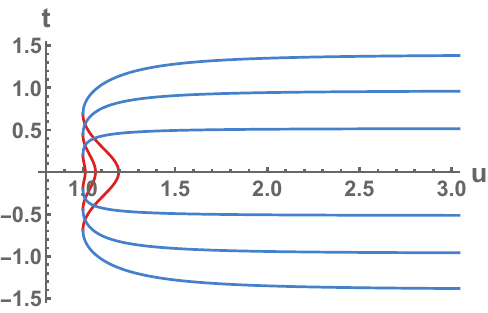}}\hspace{.5cm}
 	\subfigure[ ]{\label{fig:sol2}  \includegraphics[width=71mm]{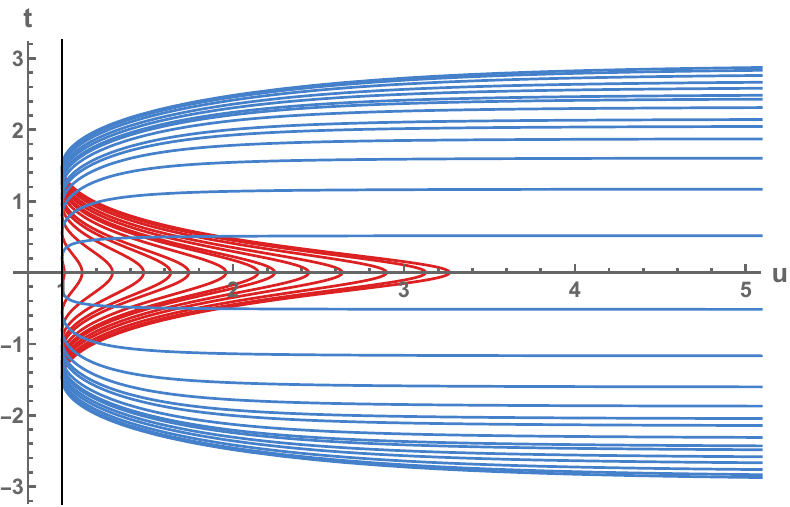}}
	\caption{The surfaces comprising the tEE. The red-colored curve is a solution of the equations of motion \cref{eqtt1} and extends from the turning point $u_0$ to $u_k$. The blue-colored curve consists of the solutions of \cref{eqtt2} and extends from the boundary to $u_k$. In the left figure, we plot only three representative solutions with a focus on the regime of $u_0$ around $u_k$. The right figure contains the line of the natural IR cut-off $u=u_k$ that is the tip of the cigar geometry and contains the curves of a wide range of turning points of $u_0$. As the turning point distances itself from the tip of the geometry, the surface increases its boundary length. For all our plots we fix $u_k=1$ and we normalize all the parameters with it, or equivalently with the fixed radius $R_0$.}
	\label{surface2}
\end{figure}

In this section, we find the surfaces comprising the tEE. The equations of motion to be solved are given by \cref{gentprime} where for convenience, following the \cref{Lorentz} we redefine
$c_0:=c/\cV_{S^4}$ to absorb the constant $\cV_{S^4}$ that appears in $V_{int}$ and we have $c_0:=i \tilde{c}_0\Rightarrow c_0^2=u_0^2\, u_k^3-u_0^5<0$:
\begin{equation}\la{eqtt1}
t'_{Im}(u):=t^\prime(u)=\pm\sqrt{\frac{c_0^2} { \prt{u^3-u_k^3}\prt{c_0^2-u^2\, u_k^3+u^5}}}~. 
\end{equation}
As described in the previous sections by looking at \cref{gentur}, we see that there is a turning point of the symmetric surface, and the surface extends from the turning point $u_0$ towards the deep IR and the tip of the geometry. The two branches of the solution in the above \cref{eqtt1} are presented by the red color in the \cref{surface,surface2}. This part of the surface is responsible for the imaginary part of the tEE as we will point out later. In the regime $u_0\simeq u_k$ the surface becomes almost flat and lies on the $u=u_k$, as depicted in \cref{surface2}. 

Moreover, for $\tilde{c}_0^2>0$ we have the equation of motion as
\begin{equation}\la{eqtt2}
t'_{Re}(u):=t^\prime(u)=\pm\sqrt{\frac{\tilde{c}_0^2} { \prt{u^3-u_k^3}\prt{\tilde{c}_0^2-u^2\, u_k^3+u^5}}}~,
\end{equation}
where $\tilde{c}_0^2=u_0^5-u_0^2\, u_k^3$. These surfaces have no turning point, the surface extends from the asymptotic boundary towards the deep IR. Therefore, their boundary points depend on the integration constant $\tilde{c}_0$, which needs to be fixed by additional information. We propose to determine the boundary length by merging smoothly the timelike and spacelike surfaces in such a way that are homologous to the boundary and have a well-defined first derivative at the merging point. 

The hypersurfaces comprising the tEE are defined by
\bea \la{sigma1}
&&\S_{Im}:=t-\int t_{Im}'(u) du + c_{Im}=0~,\\
&&\S_{Re}:=t-\int t_{Re}'(u) du + c_{Re}=0~,
\la{sigma2}
\eea
with constants $c_{Re},~c_{Im}$. The gradient vectors $g^{\m\n}\pp_\n \S$ are normal to the hypersurface $\S$ and therefore a spacelike surface $\S$ is defined via $g^{\m\n}\pp_\m \S\pp_\n \S<0$, while a timelike surface is defined via $g^{\m\n}\pp_\m \S\pp_\n \S>0$. Using the \cref{eqtt1,eqtt2} we can confirm that $\S_{Im}$ and $\S_{Re}$ are timelike and spacelike respectively in the regime they extend. Now the question that arises is how and where these two surfaces can be merged. At the tip of the geometry $u_k$ momentarily we have   
\be 
g^{\m\n}\pp_\m \S_{Re}\pp_\n \S_{Re}|_{u=u_k}=g^{\m\n}\pp_\m \S_{Im}\pp_\n \S_{Im}|_{u=u_k}=0~,
\ee
which justifies it as the unique merging point of the timelike and spacelike surfaces. 

There are two different ways of merging the surfaces which are depicted with green and blue lines in \cref{surface}. Following the above prescription we choose the blue colored surfaces, depicted also in \cref{surface2} since green surfaces violate the existence of the well defined first derivative. Moreover, the patching that generates the green surfaces leads to extra crossing points with each other. Therefore, the timelike and spacelike surfaces obtained from \cref{eqtt1,eqtt2} can be smoothly patched as depicted in \cref{surface2}. We show later that the solutions of \cref{eqtt2} contribute the real part to the tEE while the surfaces of the \cref{eqtt1} to the imaginary part.

Notice that in general analytical integrations of \cref{eqtt1,eqtt2} are intractable in our confining background as we will see in the following section. In CFTs and AdS geometries where analytic solutions for this type of surfaces exist \cite{Doi:2023zaf}, the integration constant $c$ and its sign have been specified after obtaining the analytic surface solution. In our construction for tEE in Lorentzian signature, we do that at the level of the equations of motion.

Moreover, as another side remark, we note that the construction presented above has the advantage that the merged surface is expressed as a closed homologous surface. Utilizing this fact, one can define a timelike entanglement wedge for confining background. This can be helpful in understanding the quantum information theoretic properties of the time direction. We leave this interesting issue as a future work.

\subsection{The deep IR Approximation for the Turning Point $u_0$}\label{Large_T}
First, we consider the turning points $u_0$ of the extremal surfaces to be very close to the tip $u_k$ of the cigar geometry in the confining metric described in \cref{Wmetric}. In the following, we demonstrate the computations for the extremal surfaces and the subsystem length in such a limit.

\subsubsection{Euclidean Signature, Entanglement Entropy and Simple Analytic Continuation}

To obtain the subsystem length $T_E$ in terms of the turning point, we consider the Euclidean signature of the metric in \cref{Wmetric}, and we re-express the integral in \cref{Tu0} as
\begin{align}
	T_E &=2\int_{u_0}^{\infty} du ~  u_0 \sqrt{\frac{u_0^3-u_k^3}{\left(u^3-u_k^3\right) \left(-u^2\, u_k^3+u_0^2\, u_k^3+u^5-u_0^5\right)}}~.
\end{align}
By considering the leading order terms in $u_0$ around $u_k$, the contribution from the above integration can be read as
\begin{align}\label{T_u0_eucl}
	T_E\simeq \frac{2 \sqrt{u_0-u_k} \left(\log \left(u_0-u_k\right)-\log \left(u_k\right)+\log (3)\right)}{\sqrt{3}\, u_k}~.
\end{align}
From the above expression, we consider the leading terms for which the turning point $u_0$ of the extremal surfaces can now be solved in terms of the subsystem length $T_E$ as follows
\begin{align}\label{u0_T_eucl}
	u_0 \simeq e^{2\, W_{-1}\left(-\frac{1}{4} \sqrt{3}\, T_E\, u_k\right)}+u_k~,
\end{align}
where $W$ is the Lambert function in the $W_{-1}$ branch and is defined as a solution of the equation $z = W(z)e^{W(z)}$.  From \cref{u0_T_eucl}, we note that when $T_E$ vanishes, the turning point of the connected surface $u_0$ approaches $u_k$, and is consistent with \cref{T_u0_eucl} as expected. Note that the $W$ function is the usual function that often appears in the minimal surface computation of Wilson loops and entanglement entropy in confining holographic backgrounds.

We now compute the connected and the disconnected surface in \cref{genSrenorm} for the Euclidean signature of the metric described in \cref{Wmetric}
\begin{align}\label{S_con}
	4 G_N^{(10)} \mathcal{S}_{con} &= 2 \mathcal{V}^{3}\, \cV_{S ^ 4} \int_{u_0}^{\infty} du ~\sqrt{\frac{u^7-u^4\, u_k^3}{-u^2\, u_k^3+u_0^2 \,u_k^3+u^5-u_0^5}}~,\notag\\
	4 G_N^{(10)} \mathcal{S}_{discon} &= 2 \mathcal{V}^{3}\, \cV_{S ^ 4} \int_{u_k}^{\infty} du ~ u~.
\end{align}
To derive the equations above, we have considered $d_0=5$ in \cref{genSrenorm} to express the volume as $\cV^{d_0-2}=\cV^3:=\cV^2 2 \pi R_0$ where $\cV^2$ is the area along $x_2,x_3$ directions and $\cV^3$ is the spatial volume extended in $x_2,x_3$ and the compactified dimension $y$. The connected surface contains divergence at the boundary $u=\infty$ from where we extract the finite part by subtracting $ \mathcal{S}_{discon}$ from $\mathcal{S}_{con}$ and obtain
\begin{align}\label{S_renorm_eucl}
	\frac{4 G_N^{(10)}}{ 2 \mathcal{V}^{3} \,\cV_{S ^ 4}} \mathcal{\hat{S}} &= \frac{4 G_N^{(10)}}{ 2 \mathcal{V}^{3}\, \cV_{S ^ 4}} \left( \mathcal{S}_{con}-\mathcal{S}_{discon}\right)\simeq-u_k (u_0-u_k) \log (u_0-u_k)~,
\end{align}
for the leading contributions in $u_0 \sim u_k$ regime for the integrals of \cref{S_con}. Now, utilizing \cref{u0_T_eucl}, we can re-express $\mathcal{\hat{S}}$ in \cref{S_renorm_eucl} in terms of the subsystem length $T_E$ as
\begin{align}\label{S_eucl}
	\frac{4 G_N^{(10)}}{ 2 \mathcal{V}^{3}\, \cV_{S ^ 4}} \mathcal{\hat{S}} &\simeq -2 u_k \,e^{2\, W_{-1}\left(-\frac{1}{4}\sqrt{3}\, T_E\, u_k\right)} W_{-1}\left(-\frac{1}{4}
   \sqrt{3}\, T_E\, u_k\right) ~.
\end{align}
The above expression yields that in the limit $u_0\to u_k$, $\mathcal{\hat{S}}$ vanishes in the confining background which implies that the connected surfaces approach the disconnected surface in the regime $u_0\sim u_k$.

For completeness we present a naive analytic continuation $T_E \rightarrow i\, T$ in \cref{S_eucl}, which gives for the analytically continued  entanglement entropy 
\begin{align}\label{S_T_wick}
	\frac{4 G_N^{(10)}}{ 2 \mathcal{V}^{3}\, \cV_{S ^ 4}} \mathcal{\hat{S}}^T &\simeq -2 u_k\, e^{2\, W_{-1}\left(-\frac{\sqrt{3}}{4}\, i\, T\, u_k\right)} W_{-1}\left(-\frac{\sqrt{3}}{4}\,
    i\, T\, u_k\right) ~.
\end{align}
Notice that in this limit we are in the unfavorable higher blanch of the EE, as can be seen in the \cref{ee_num} in the later numerical sections. Nevertheless, the analytic expressions give insights into the behavior of the analytically continued EE.

\subsubsection{Timelike Entanglement Entropy in Lorentzian Signature}
Let us consider the confining background and compute the subsystem length and the timelike entanglement entropy by considering the leading terms in $u_0$ around the tip $u_k$ of the cigar geometry.

To obtain the subsystem length, we integrate the integral of \cref{eqtt1}  and \cref{eqtt2}. The T-length of the surface that starts from the boundary is
\begin{align}\la{t1}
	T_{Re} &=2\int_{u_k}^{\infty} du ~ \sqrt{\frac{\tilde{c}_0^2} { \prt{u^3-u_k^3}\prt{\tilde{c}_0^2-u^2\, u_k^3+u^5}}}~, 
\end{align}
where the multiplication by $2$ is because we have two symmetric surfaces. The surfaces always extend from the boundary to the tip of the geometry. Nevertheless, we have $c_0= i\tilde{c}_0$ which from \cref{gentur} gives
$c_0^2=-u_0^2\prt{u_0^3-u_k^3}$ 
where $u_0$ is the turning point of the symmetric bulk surface that contributes to the imaginary part of the tEE. 
When $\tilde{c}_0\rightarrow0$ the integral $T_{Re}\rightarrow 0$. In order to understand analytically the behavior of the surface we make the following assumption, to integrate $T_{Re}$ from $u_0\simeq u_k$ to $\infty$ where $|\tilde{c}_0|\ll 1$.
For the leading contributions we obtain
\be\la{Tu0_Re_high_temp}
    T_{Re}\simeq-\frac{2 \sqrt{u_0-u_k} \log \left(u_0-u_k\right)}{\sqrt{3}\, u_k}~.
\ee

Next, let us consider the bulk surface in the regime near the tip where the integral is squeezed by its endpoints since the bounds tend towards each other
\begin{align}\la{t2}
T_{Im} &=2\int_{u_k}^{u_0} du ~  \sqrt{\frac{c_0^2} {\prt{u^3-u_k^3}\prt{c_0^2-u^2\, u_k^3+u^5}}}~.
\end{align}
The integrand diverges at $u_k$ and therefore the near $u_k$ convergence of the integral is not linear. We can approximate the convergence to zero as
\be\la{Tu0_Im_high_temp}
    T_{Im}\simeq\frac{14 \sqrt{u_0-u_k}}{3 \sqrt{3}\, u_k}~.
\ee
 
The total subsystem including the contributions from $\tilde{c}_0^2>0$ and $c_0^2<0$ solutions can be expressed as a function of $u_0$ from \cref{Tu0_Re_high_temp,Tu0_Im_high_temp} 
\begin{align}\la{T_u0_Low_temp}
T\left(u_0\right)&=T_{Re}\left(u_0\right)+T_{Im}\left(u_0\right)
    \simeq \frac{2 \sqrt{u_0-u_k} \left(7-3 \log \left(u_0-u_k\right)\right)}{3 \sqrt{3} \,u_k}~.
\end{align}
Inverting the above expression, we can solve for the turning point $u_0$ as a function of the total subsystem $T$ as
\be\la{T_tot_u0_high_temp}
    u_0\simeq e^{\frac{1}{3} \left(7+6\, W_{-1}\left(-\frac{\sqrt{3}\, T \,u_k}{4\, e^{7/6}}\right)\right)}+u_k~,
\ee
where $W$ is the Lambert function in the $W_{-1}$ branch.

The real part of the extremal surfaces in \cref{genSint2} and the disconnected surface in \cref{genSdis} are given by
\begin{align}\label{S_Re}
4 G_N^{(10)} \mathcal{S}^T_{\tilde{c}_0} &= 2 \mathcal{V}^{3}\, \cV_{S ^ 4}\int_{u_k}^{\infty} du ~u^2\sqrt{\frac{u^3- u_k^3}{u^5-u_0^5-u^2\, u_k^3+u_0^2\, u_k^3}}~,\notag\\
	4 G_N^{(10)} \mathcal{S}^T_{discon} &= 2 \mathcal{V}^{3}\, \cV_{S ^ 4}\int_{u_k}^{\infty} du ~ u~.
\end{align}
We remove the divergence in the area of the extremal surfaces at the boundary $u=\infty$ by subtracting $ \mathcal{S}^T_{discon}$ from $\mathcal{S}^T_{\tilde{c}_0}$. Furthermore, exploiting the fact that the turning point is very close to the tip of the geometry, we consider the integration from $u_0\simeq u_k$ to infinity as
\begin{align}\label{S_real_highT}
	\frac{4 G_N^{(10)}}{ \mathcal{V}^{3}\, \cV_{S ^ 4}} \mathcal{\hat{S}}^T_{Re} &= \frac{4 G_N^{(10)}}{ \mathcal{V}^{3}\,  \cV_{S ^ 4}} \left( \mathcal{S}^T_{\tilde{c}_0}-\mathcal{S}^T_{discon}\right)
	\simeq -u_k\left(u_0-u_k\right) \log ( u_0-u_k )~.
\end{align}
Utilizing \cref{T_tot_u0_high_temp}, we can re-express the timelike entanglement entropy in \cref{S_real_highT} in terms of the total subsystem length $T$ as
\begin{align}
	\frac{4 G_N^{(10)}}{ \mathcal{V}^{3}\, \cV_{S^4}} \mathcal{\hat{S}}^T_{Re} &\simeq -e^{\frac{1}{3} \left(7+6\, W_{-1}\left(-\frac{\sqrt{3}\, T\, u_k}{4\, e^{7/6}}\right)\right)} u_k \log e^{\frac{1}{3} \left(7+6\, W_{-1}\left(-\frac{\sqrt{3}\, T\, u_k}{4\, e^{7/6}}\right)\right)} ~.
\end{align}
The above expression shows that in the limit $u_0\to u_k$, the real part of $\mathcal{\hat{S}}^T$ vanishes in the confining background which implies that the real part of the extremal surfaces asymptotically approaches the disconnected surfaces in the regime $u_0\sim u_k$.

Similar to the real part described above, we can obtain the imaginary part of the extremal surfaces in \cref{genSint2} using $c_0^2<0$,
\begin{align}\label{S_Im_highT}
		4 G_N^{(10)} \mathcal{\hat{S}}^T_{Im}&= 4 G_N^{(10)} \mathcal{S}^T_{c_0}
		 =2\, i\, \mathcal{V}^{3}\,  \cV_{S^4} \int_{u_k}^{u_0} du ~u^2 \sqrt{\frac{u^3- u_k^3}{u^5+u_0^5-u^2\, u_k^3-u_0^2\, u_k^3}}~.
\end{align}
The surface is squeezed between $u_0$ and $u_k$, where we can approximate it as
\begin{align}
    \frac{4 G_N^{(10)}}{ \mathcal{V}^{3}\, \cV_{S^4}} \mathcal{\hat{S}}^T_{Im} &\simeq i\,\ff{4}{3}\,  u_k (u_0 - u_k)\simeq i\,\ff{4}{3}\, u_k\, e^{\frac{1}{3} \left(7+6\, W_{-1}\left(-\frac{\sqrt{3}\, T\, u_k}{4\, e^{7/6}}\right)\right)}~,
\end{align}
where for the second equality we have used \cref{T_tot_u0_high_temp}.
As earlier, the above expression justifies that in the limit $u_0\to u_k$, the imaginary part of $\mathcal{\hat{S}}^T$ vanishes in the confining background. This is expected since this surface is squeezed between $u_0$ and $u_k$.  Notice that for our analytic approach, we have considered several assumptions. Nevertheless, it serves mainly as a successful approximation to understand the limits of our surfaces and when they diminish or not.

\subsection{Near Boundary Behavior}\label{u0_greaterthan_uk}
We now consider the limit where the turning point $u_0$ of the extremal timelike surface stays close to the boundary $u=\infty$ for the confining background described in \cref{Wmetric}. In this section, we perform binomial expansions for the approximations.

 \subsubsection{Euclidean Signature, Entanglement Entropy and simple Analytic Continuation}
We first consider the Euclidean signature of the confining metric described in \cref{Wmetric} and compute the subsystem length $T_E$ from the integration expressed in \cref{Tu0} in the limit $u_0 \gg u_k$. Utilizing the metric elements given in \cref{Wmetric}, the corresponding integral in \cref{Tu0} reads as
\begin{align}\label{T_E_all}
	T_E &=2\int_{u_0}^{\infty} du ~  u_0 \sqrt{\frac{u_0^3-u_k^3}{\left(u^3-u_k^3\right) \left(-u^2\, u_k^3+u_0^2\, u_k^3+u^5-u_0^5\right)}}~.
\end{align}
Now we consider the change of variables
\be\la{var}
x=\frac{u_0}{u}, \quad p=\frac{u_k}{u_0}~,\quad y= p\;x =\frac{u_k}{u}~,
\ee
to express \cref{T_E_all} as
\begin{align}\label{Te_b}
    T_E &=2\int_{0}^{1} dx~ \frac{x^2 \sqrt{1-p^3}}{\sqrt{u_0} \sqrt{\left(1-f(x,y) \right) \left(1-y^3\right)}} ~,
\end{align}
where $f(x,y)$ is given by
\begin{align}\label{fxy}
    f(x,y)=x^5+y^3-x^2 y^3~.
\end{align}
The function in the above equation always satisfies the inequality $f(x,y)\leq 1$ in the regime $u \in [u_0,\infty]$.
We perform binomial expansions for $f(x,y)\leq 1$ and $y\leq 1$ and rewrite \cref{Te_b}  as a complete sum
\begin{align}
    T_E =\frac{2\sqrt{1-p^3}}{\sqrt{u_0}}\int_{0}^{1} dx \sum _{k,l=0}^{\infty } (-1)^{k+l} p^{3 k} a_k\, b_l\,
    x^{3k+3l+2} \left( x^2+p^3\left(1-x^2\right) \right)^l,
\end{align}
where
\begin{equation}
    a_k=\frac{\sqrt{\pi }}{2\, \Gamma \left(\frac{3}{2}-k\right) \Gamma (k+1)}~, \qquad \qquad b_l=\frac{\sqrt{\pi }}{\Gamma \left(\frac{1}{2}-l\right) \Gamma (l+1)}~.
\end{equation}
From the above equation, we can extract the terms up to the first order correction in $u_0$ as
\begin{align}\label{Tu0gguk}
    T_E &\simeq \frac{1}{\sqrt{u_0}}\left(\frac{1}{\alpha }-\frac{u_k^3}{u_0^3}\,\frac{\beta}{2}\right)~,
\end{align}
where $\alpha$ and $\beta$ are numerical constants given by
\be\la{alpha_beta}
    \alpha=\frac{\Gamma \left(\frac{1}{10}\right)}{4 \sqrt{\pi}\,  \Gamma \left(\frac{3}{5}\right)}~, \qquad\qquad
    \beta= \frac{4 \sqrt{\pi }\, \Gamma \left(\frac{1}{5}\right)}{35\, \Gamma \left(\frac{7}{10}\right)}~.
\ee
Note that the subsystem length $T_E$ vanishes at the asymptotic boundary $u_0\to\infty$. 
The turning point of the connected surface $u=u_0$ can now be expressed in terms of the subsystem length $T_E$ as follows
\begin{align}\label{u0T_u0gguk}
    u_0\simeq\frac{1}{\alpha^2 \, T_E^2 \left(1+\alpha^7 \beta \, T_E^6 \, u_k^3\right)}~,
\end{align}

We proceed with the computation of the difference between the areas of the connected and the disconnected surfaces following \cref{genSrenorm}. The connected and the disconnected surfaces for the confining background can be expressed as described in \cref{S_con}. Utilizing the change of variables in \cref{var}, we rewrite the expressions in \cref{S_con} as follows
\begin{align}
    4 G_N^{(10)} \mathcal{S}_{con} &= 2 \mathcal{V}^{3}\, \cV_{S ^ 4}\, u_0^2 \int_{0}^{1} dx \frac{\sqrt{1-y^3}}{x^3 \sqrt{1-f(x,y)}}~,\notag\\
	4 G_N^{(10)} \mathcal{S}_{discon} &= 2 \mathcal{V}^{3}\, \cV_{S ^ 4}\, u_0^2 \int_{0}^{1/p} \frac{dx}{x^3}~.
\end{align}
In the above expression, $f(x,y)$ is given by \cref{fxy} which again satisfies the inequality $f(x,y)\leq 1$ in the parameter regime $u \in [u_0,\infty]$. We may now rewrite the integral of $\mathcal{S}_{con}$ as a complete sum by utilizing binomial expansions:
\begin{align}
    4 G_N^{(10)}& \mathcal{S}_{con} = 2 \mathcal{V}^{3}\, \cV_{S ^ 4}\, u_0^2 \int_{0}^{1} dx \sum _{k,l=0}^{\infty } (-1)^{k+l}p^{3 k} a_k\,b_l\,x^{3k+3l-3} \left( x^2+p^3\left(1-x^2\right) \right)^l.
\end{align}
The above expression indicates a divergence of the area of the connected surface at the boundary $u_0=\infty$ ($x=0$). We therefore define a finite quantity by subtracting the disconnected surface from the connected surface which in the limit of $u_0\gg u_k$ can be obtained in the leading order of $u_0$ as
\begin{align}
    \frac{4 G_N^{(10)}}{ 2 \mathcal{V}^{3}\, \cV_{S ^ 4}} \mathcal{\hat{S}} &= \frac{4 G_N^{(10)}}{ 2 \mathcal{V}^{3}\, \cV_{S ^ 4}} \left( \mathcal{S}_{con}-\mathcal{S}_{discon}\right) 
    \simeq u_k^2- \frac{u_0^2}{4\alpha}~,
\end{align}
where the numerical constant $\alpha$ in the above expression is given by \cref{alpha_beta}.
Note that for $u_0\gg u_k$, the connected surface $\mathcal{S}_{con}$ always dominates over the disconnected surface $\mathcal{S}_{discon}$. 
Utilizing \cref{u0T_u0gguk}, we can re-write $\mathcal{\hat{S}}$ in terms of the subsystem length
\begin{align}
    \frac{4 G_N^{(10)}}{ 2 \mathcal{V}^{3}\, \cV_{S ^ 4}} \mathcal{\hat{S}} & \simeq u_k^2-\frac{1}{4\, \alpha ^5 \,T_E^4 \left(1+\alpha^7 \beta \, T_E^6 \, u_k^3\right)^2}~.
\end{align}
We can now perform a naive Wick rotation on the subsystem length $T_E\rightarrow i\, T$ to obtain the analytically continued entanglement entropy
\begin{align}
    \frac{4 G_N^{(10)}}{ 2 \mathcal{V}^{3}\, \cV_{S ^ 4}} \mathcal{\hat{S}}^T &\simeq u_k^2-\frac{1}{4\, \alpha ^5 \,T^4 \left(1-\alpha^7 \beta \, T^6 \, u_k^3\right)^2}~.
\end{align}
The above result indicates that there is a critical length $T=1/\left(\alpha^7 \beta u_k^3\right)^{1/6}$ where $\mathcal{\hat{S}}^T$ becomes infinity.

\subsubsection{Timelike Entanglement Entropy in Lorentzian Signature}
In this subsection, we consider the Lorentzian signature of the confining background described in \cref{Wmetric} and focus on the limit $u_0\gg u_k$ to compute the subsystem lengths and the area of the corresponding extremal surfaces for the two solutions $c_0$ and $\tilde{c}_0$ described in \cref{TEELor}.

We obtain the subsystem lengths by integrating the expressions in \cref{eqtt1} and \cref{eqtt2} with respect to the bulk direction $u$. For the surface corresponding to $\tilde{c}_0^2>0$ that extends from the asymptotic boundary $u=\infty$ to the tip $u=u_k$ of the cigar geometry, the corresponding subsystem length is given by the integral in \cref{t1}. We now utilize the change of variables in \cref{var}  to re-express the integral of the subsystem length as
\begin{align}\label{Tl_b}
    T_{Re}= 2\frac{\sqrt{1-p^3}}{\sqrt{u_0}} \left(\int_{1}^{1/p} \frac{dx}{ \sqrt{x} \sqrt{\left(1-y^3\right) \left(f_1(x,y)+1\right)}} + \int_0^1 \frac{x^2 dx}{ \sqrt{\left(1-y^3\right) \left(f_2(x,y)+1\right)}}\right)~.
\end{align}
where the functions 
\begin{align}
    f_1(x,y)&=\frac{1-y^3+x^2 y^3}{x^5}~,\label{f1xy}\\
    f_2(x,y)&=x^5-y^3-x^2 y^3~,\label{f2xy}
\end{align}
satisfy the inequalities $f_{1,2}(x,y)\leq 1$ in the parameter regimes $u\in [u_k,u_0]$ and $u\in[u_0,\infty]$ respectively. We rewrite \cref{Tl_b} as a complete sum by utilizing binomial expansions for $y\leq1$ and $f_{1,2}(x,y)\leq1$ as
\begin{align}
    T_{Re} &= 2\frac{\sqrt{1-p^3}}{\sqrt{u_0}} \int_{1}^{1/p} dx \sum _{k,l=0}^{\infty } (-1)^k p^{3 k} (1-2k)\,a_k\,b_l\, x^{3 k-5l-\frac{1}{2}} \left(1-p^3 x^3\left( 1-x^2\right)\right)^l\nn\\
    &+ 2\frac{\sqrt{1-p^3}}{\sqrt{u_0}} \int_0^1 dx \sum _{k,l=0}^{\infty } (-1)^k p^{3 k} (1-2k)\,a_k\,b_l\, x^{3k+3l+2} \left( x^2-p^3\left(1+x^2\right) \right)^l~.
\end{align}
Considering the leading order terms in $u_0$, we may compute the subsystem length $T_{Re}$ in the limit $u_0\gg u_k$ for the real part of the extremal surfaces as
\begin{align}\label{Tu0_Re_u0gguk_app}
    T_{Re}\simeq \frac{1}{\sqrt{u_k}}\frac{2 \sqrt{3 \pi }\, \Gamma \left(\frac{4}{3}\right)}{ \Gamma \left(\frac{5}{6}\right)} -\frac{1}{\sqrt{u_0}}\frac{4\, \Gamma \left(\frac{3}{5}\right) \Gamma \left(\frac{9}{10}\right)}{\sqrt{\pi } }~.
\end{align}

The imaginary part of the extremal surfaces for the solution $c_0^2<0$ extends from the turning point $u=u_0$ to the tip of the cigar geometry $u=u_k$. The corresponding subsystem length is given by the integral described in \cref{t2}. We can now re-express the subsystem length $T_{Im}$ for the imaginary surface with the change of variables in \cref{var} as
\begin{align}
    T_{Im} = \frac{2 \sqrt{1-p^3}}{\sqrt{u_0}} \int^{1/p}_1 \frac{dx}{\sqrt{x} \sqrt{\left(1-y^3\right) \left(1-f_1(x,y)\right)}}~,
\end{align}
where the expression for $f_1(x,y)$ is given by \cref{f1xy}. As earlier we perform binomial expansions for $y\leq1$ and $f_1(x,y)\leq1$ to obtain $T_{Im}$ as a complete sum
\begin{align}
    T_{Im} &= \frac{2 \sqrt{1-p^3}}{\sqrt{u_0}} \int^{1/p}_1 \sum _{k,l=0}^{\infty } (-1)^{k+l} p^{3 k} (1-2k)\,a_k\,b_l\, x^{3k-5l-\frac{1}{2}} \left( x^2-p^3\left(1+x^2\right) \right)^l~.
\end{align}
Utilizing the above expression, we can now extract the leading order terms in $u_0$ in the limit $u_0\gg u_k$ as
\begin{align}\label{Tu0_Im_u0gguk}
    T_{Im}\simeq \frac{1}{\sqrt{u_k}}\frac{2 \sqrt{3 \pi }\, \Gamma \left(\frac{4}{3}\right)}{ \Gamma \left(\frac{5}{6}\right)} - \frac{1}{\sqrt{u_0}}\frac{4 \sqrt{\pi }\, \Gamma \left(\frac{9}{10}\right)}{ \Gamma \left(\frac{2}{5}\right)}~.
\end{align}
Note that in the limit $u_0\gg u_k$, the leading terms containing $\frac{1}{\sqrt{u_k}}$ in the expressions for $T_{Re}$ and $T_{Im}$ are equal which implies that the two subsystems approach each other. More importantly, notice that there are maximum lengths of $T_{Im}$ and $T_{Re}$.

The total subsystem $T$ can now be expressed including the contributions from the solutions $\tilde{c}_0^2>0$ and $c_0^2<0$ as
\begin{align}\la{u0_T_tot}
T\left(u_0\right)&=T_{Re}\left(u_0\right)+T_{Im}\left(u_0\right)
    =\frac{\gamma }{\sqrt{u_k}}-\frac{1}{\sqrt{u_0}}\frac{\delta }{\alpha  }~,
\end{align}
where $\alpha$ is given by \cref{alpha_beta} and $\gamma$, $\delta$ are other numerical constants
\be\la{gamma_delta}
    \gamma=\frac{4 \sqrt{3 \pi }\, \Gamma \left(\frac{4}{3}\right)}{\Gamma \left(\frac{5}{6}\right)}~, \qquad\qquad \delta=1+\sqrt{5}+\sqrt{2 \sqrt{5}+5}~.
\ee
Note that in the limit $u_0 \gg u_k$, the leading contribution to the subsystem length $T$ comes from the first term in \cref{u0_T_tot} such that $T$ has a maximum value of $T_{crit}=\gamma/\sqrt{u_k}$.
From the above expression, we can solve for the turning point $u_0$ as a function of the total subsystem length $T$ as
\be\la{T_tot_u0_low_temp}
 u_0\simeq \frac{\delta ^2 u_k}{\alpha ^2 \left(\gamma -T \sqrt{u_k}\right)^2}~.
\ee
From \cref{gamma_delta}, we notice that at the maximum value of the subsystem $T\simeq T_{crit}=\gamma/\sqrt{u_k}$, the turning point moves towards the asymptotic boundary $u_0 \rightarrow \infty$.

Now we proceed to compute the real and imaginary parts of the area of the extremal surfaces for the two solutions corresponding to $\tilde{c}_0^2>0$ and $c_0^2<0$ respectively. The area of the extremal surface for $\tilde{c}_0^2>0$ and the disconnected surface are given by the expressions in \cref{S_Re}. As earlier we utilize the change of variables described in \cref{var} to re-express the corresponding areas of the extremal surfaces as
\begin{align}
    4 G_N^{(10)} \mathcal{S}^T_{\tilde{c}_0} &= 2 \mathcal{V}^{3}\, \cV_{S ^ 4}\, u_0^2 \left( \int_{1}^{1/p} \frac{\sqrt{1-y^3}}{x^{11/2}\sqrt{f_3(x,y)+1}} + \int_{0}^{1} dx \frac{\sqrt{1-y^3}}{x^3\sqrt{f_2(x,y)+1}} \right), \notag\\
    4 G_N^{(10)} \mathcal{S}_{discon} &= 2 \mathcal{V}^{3}\, \cV_{S ^ 4}\, u_0^2 \int_{0}^{1/p} \frac{dx}{x^3}~,
\end{align}
where the function $f_2(x, y)$ is given by \cref{f2xy} and the function $f_3(x,y)$ is described by
\begin{align}\label{f3xy}
    f_3(x,y)=\frac{1-y^3-x^2 y^3}{x^5}~,
\end{align}
which satisfies the inequality $f_3(x,y)\leq 1$ in the parameter regime $u\in [u_k,u_0]$. Once again, we perform binomial expansions for $y\leq 1$ and $f_{2,3}(x,y)\leq 1$ to obtain the above integral as complete sum
\begin{align}
    4G_N^{(10)} \mathcal{S}^T_{\tilde{c}_0}
    &= 2 \mathcal{V}^{3}\, \cV_{S ^ 4}\, u_0^2 \int_{1}^{1/p}  dx \sum _{k,l=0}^{\infty } (-1)^k p^{3 k} a_k\,b_l\,x^{3 k-5l-\frac{11}{2}} \left(1-p^3 x^3 \left(1+x^2 \right)\right)^l\nn\\
    &+ 2 \mathcal{V}^{3}\, \cV_{S ^ 4}\, u_0^2 \int_{0}^{1}  dx \sum _{k,l=0}^{\infty } (-1)^k p^{3 k}a_k\,b_l\, x^{3k+3l-3}\left( x^2-p^3\left(1+x^2\right) \right)^l~.
\end{align}
The real surface expressed above contains divergence at the asymptotic boundary $u=\infty$. Thus, we extract the finite part by subtracting $ \mathcal{S}^T_{discon}$ from $\mathcal{S}^T_{\tilde{c}_0}$, which in the limit $u_0\gg u_k$ can be obtained in the leading order of $u_0$ as
\begin{align}\label{S_renorm_highT}
    \frac{4 G_N^{(10)}}{ \mathcal{V}^{3}\, \cV_{S ^ 4}} \mathcal{\hat{S}}^T_{Re} &= \frac{4 G_N^{(10)}}{ \mathcal{V}^{3}\,  \cV_{S ^ 4}} \left( \mathcal{S}^T_{\tilde{c}_0}-\mathcal{S}^T_{discon}\right)\simeq u_k^2- u_0^2 \frac{\Gamma \left(\frac{3}{5}\right) \Gamma \left(\frac{9}{10}\right)}{\sqrt{\pi } }= u_k^2- u_0^2\frac{\left(\sqrt{5}+1\right)}{4 \alpha }~.
\end{align}
where the constant $\alpha$ is given by \cref{alpha_beta}.
Utilizing \cref{T_tot_u0_low_temp}, we rewrite the above expression in terms of the total subsystem length $T$
\begin{align}\label{S_renorm_highT2}
    &\frac{4 G_N^{(10)}}{ \mathcal{V}^{3}\, \cV_{S ^ 4}} \mathcal{\hat{S}}^T_{Re} \simeq u_k^2\left[1 -\frac{\left(\sqrt{5}+1\right) \delta ^4 }{4 \alpha ^5 \left(\gamma -T \sqrt{u_k}\right)^4}\right]~,
\end{align}
where the constants $\gamma$ and $\delta$ are given by \cref{gamma_delta}. We note from the above expression that $\mathcal{\hat{S}}^T_{Re}$ diverges at the maximum value of the subsystem length $T\simeq T_{crit}=\gamma/\sqrt{u_k}$.

Lastly, the imaginary part of the extremal surfaces for the solution $c_0^2<0$ is given by \cref{S_Im_highT} which can be re-expressed utilizing the change of variables described in \cref{var} as 
\begin{align}
		4 G_N^{(10)} \mathcal{\hat{S}}^T_{Im}&= 4 G_N^{(10)} \mathcal{S}^T_{c_0}
		 =2 i \mathcal{V}^{3}\,  \cV_{S^4}\, u_0^2 \int_{1}^{1/p} \frac{\sqrt{1-y^3}}{x^{11/2}\sqrt{1-f_1(x,y)}}~,
\end{align}
where the function $f_1(x,y)$ is given by \cref{f1xy}. Once again, we perform binomial expansions for $y\leq 1$ and $f_1(x,y)\leq 1$ to obtain the complete sum for the above integral
\begin{align}
    &4 G_N^{(10)} \mathcal{\hat{S}}^T_{Im}
    = 2i \mathcal{V}^{3}\,  \cV_{S^4}\, u_0^2 \int_{1}^{1/p} \sum _{k,l=0}^{\infty } (-1)^{k+l} p^{3 k}a_k\,b_l\, x^{3k-5l-\frac{11}{2}} \left(1-p^3 x^3\left(1-x^2\right)\right)^l~.
\end{align}
In the limit $u_0\gg u_k$, we  obtain
\begin{align}\label{S_Im_lowT}
    \frac{4 G_N^{(10)}}{\mathcal{V}^{3}\,  \cV_{S^4}} \mathcal{\hat{S}}^T_{Im}
    &\simeq i\,u_0^2 \frac{\sqrt{\pi }\, \Gamma \left(\frac{9}{10}\right)}{\Gamma \left(\frac{2}{5}\right)} - i\frac{u_k^3}{u_0 }\frac{2 \sqrt{\pi }\, \Gamma \left(\frac{13}{10}\right)}{3\, \Gamma \left(\frac{4}{5}\right)} 
    = i\,u_0^2 \frac{ \sqrt{2 \sqrt{5}+5}~}{4 \alpha} - i\frac{u_k^3}{u_0}\rho ~,
\end{align}
where the constant $\rho$ is given by
\be\la{}
    \rho=\frac{2 \sqrt{\pi }\, \Gamma \left(\frac{13}{10}\right)}{3\, \Gamma \left(\frac{4}{5}\right)}~.
\ee
Notice that from \cref{S_renorm_highT,S_Im_lowT},  in the near boundary region $u_0\to\infty$, the ratio of $\mathcal{\hat{S}}^T_{Re}$ and $\mathcal{\hat{S}}^T_{Im}$ approaches the value 1.05, which will be also confirmed numerically in the next section.  Finally, we express \cref{S_Im_lowT} in terms of the total subsystem length $T$ using \cref{T_tot_u0_low_temp} as,
\begin{align}
    &\frac{4 G_N^{(10)}}{\mathcal{V}^{3}\,  \cV_{S^4}} \mathcal{\hat{S}}^T_{Im}\simeq  i\, u_k^2\Bigg[\frac{ \sqrt{2 \sqrt{5}+5} \,\delta ^4}{4\, \alpha ^5 \left(\gamma -T \sqrt{u_k}\right)^4}-\frac{ \alpha ^2 \rho \, \left(\gamma -T \sqrt{u_k}\right){}^2}{\delta ^2}\Bigg]~.
\end{align}
Similar to the real part, we note that $\mathcal{\hat{S}}^T_{Im}$ diverges at the maximum subsystem length $T\simeq T_{crit}=\gamma/\sqrt{u_k}$, since the proper length of the surface diverges at this limit.

\section{Numerical Computation of the Timelike Entanglement Entropy}
\label{numerical_timilike}

In the previous section, we have discussed the analytic computations for tEE in the limits $u_0\to u_k$ and $u_0\gg u_k$. However, for the intermediate values of the parameter $u_0$ analytic computations are in general intractable. In this section, we numerically compute the tEE in the Lorentzian confining background. We also compute the EE for a subsystem considered in the Euclidean time direction on a constant space slice. The numerical results can be presented as a substantiation of the analytic computations discussed in \cref{analytic_timelike}.

\begin{figure}[ht]
	\centering
 \subfigure[  $T_{Re}$ (red-dotted) and $T_{Im}$ (blue-dashed) $vs$ $u_0$. ]{\label{T_num}\includegraphics[width=73mm]{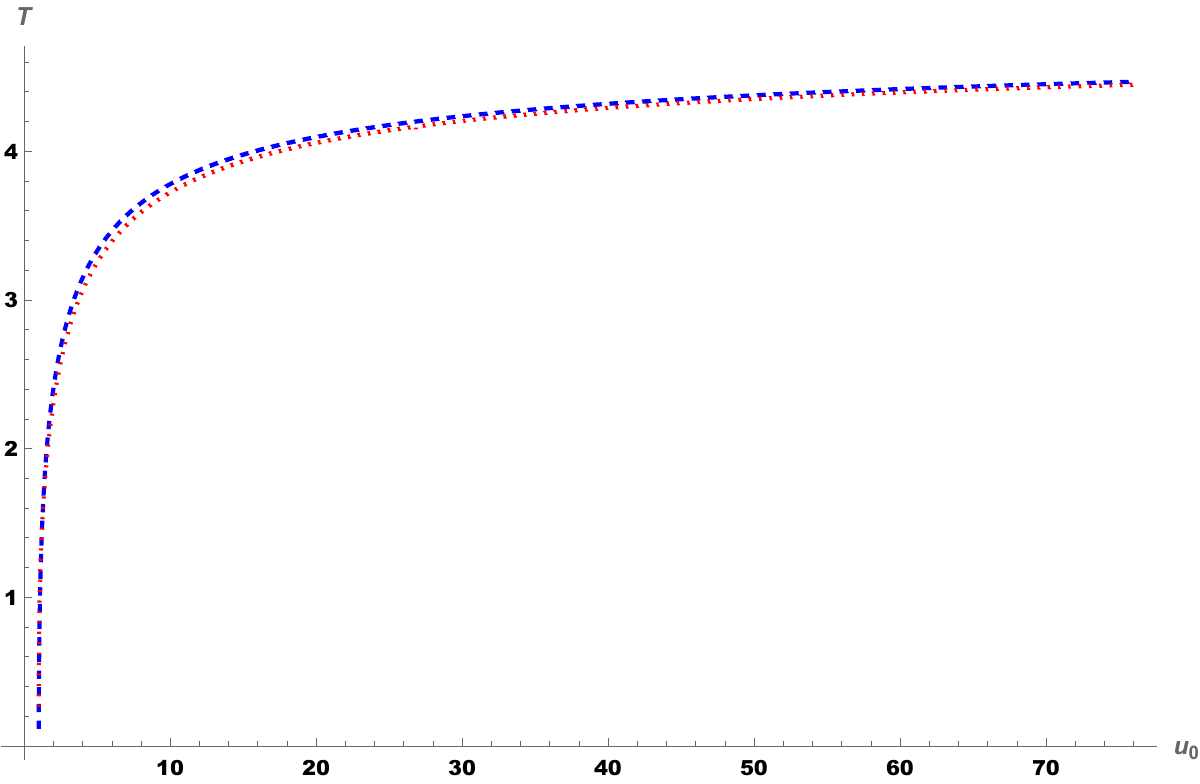}}\hspace{.5cm}
 \subfigure[ $\frac{T_{Re}}{T_{Im}}$ $vs$ $u_0$.]{\label{d_T_num}\includegraphics[width=73mm]{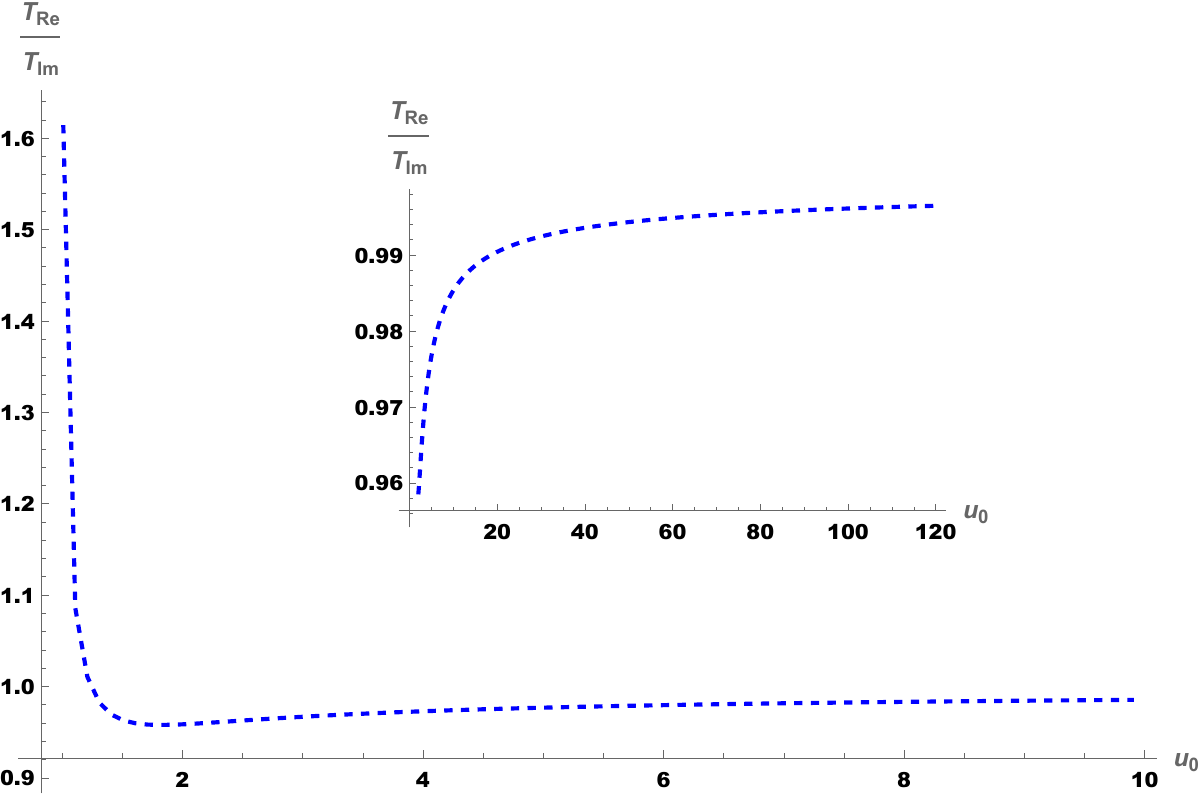}}
	\caption{The subsystem lengths as a function of $u_0$. (a) The subsystem lengths $T_{Im}$ and $T_{Re}$ corresponding to the solutions described in \cref{eqtt1,eqtt2} are plotted with increasing $u_0$. Notice the maximum value that both lengths have which is in exact agreement with the analytic computations \eq{u0_T_tot}. Both the subsystem lengths increase faster when the turning point is deep inside the bulk. (b) The ratio $\frac{T_{Re}}{T_{Im}}$ versus $u_0$. In the region where $u_0$ is close to $u_k$, it is above the unit. With increasing $u_0$ the ratio drops below the unit indicating $T_{Im}>T_{Re}$. In $u_0\gg u_k$, we observe $T_{Im}\approx T_{Re}$.
 }
	\label{num1}
\end{figure}

As explained in \cref{Lorentz}, the solution of the equation of motion indicates the possibilities of two distinct extremal surfaces. One of the surfaces extends from $u=u_k$ to the turning point $u=u_0$ which yields an imaginary value of the area. The other solution corresponds to a pair of disconnected surfaces at $t=\pm \ff{T}{2}$ which are stretched from $u=\infty$ to $u=u_k$. The area of this specific pair of surfaces contributes real values but contains UV divergences. Following the scheme explained in \cref{Lorentz} and using \cref{genSint2,genSdis}, we write $\mathcal{\hat S}^T$ for the numerical computation similar to \cref{TEELor2}  as
\begin{align}\label{areagen}
	\mathcal{\hat S}^T(u_0)&=\mathcal{\hat S}^T_{Re}(u_0)+\mathcal{\hat S}^T_{Im}(u_0)\nn\\&=\frac{\cV^{d_0-2}}{2G_N^{d+1}} \Bigg[\int_{u_k}^{\frac{1}{\epsilon }}du~V_{int}~\sqrt{g_{uu} ~g_{yy}}~g_{xx}^{\frac{d_0-3}{2}} \left(\sqrt{\frac{V^2_{int}~g_{tt}~g_{yy}~g_{xx}^{d_0-3}}{V^2_{int}~g_{tt}~g_{yy}~g_{xx}^{d_0-3}-\tilde{c}^2}}-1\right) \nn \\
&\qquad\qquad\qquad~+\int_{u_k}^{u_0}du~V_{int}~\sqrt{g_{uu}~g_{yy}}~g_{xx}^{\frac{d_0-3}{2}} \left(\sqrt{\frac{V^2_{int}~g_{tt}~g_{yy}~g_{xx}^{d_0-3}}{V^2_{int}~g_{tt}~g_{yy}~g_{xx}^{d_0-3}-c^2}}\right) \,\Bigg]~.
\end{align}
In the above equation, the constants $\tilde{c}^2=-V_{int}^2~g_{tt}~g_{yy}~g_{xx}^{d-3}|_{u=u_0}$ and $c^2=V_{int}^2~g_{tt}~g_{yy}~g_{xx}^{d-3}|_{u=u_0}$ have positive and negative values respectively, $d_0=5$  as in the rest of the text and $\epsilon\ll1$. For the numerical plots, we scale tEE with the constant $\frac{\cV^{d_0-2}}{4G_N^{d+1}}$. The length of the subsystem is computed following \cref{t1,t2} as
\begin{align}\label{tgen}
T(u_0)&=T_{Re}(u_0)+T_{Im}(u_0)\nn\\
 &=2\int_{u_k}^{\frac{1}{\epsilon}}\frac{\tilde{c}~(g_{uu}/g_{tt})^\frac{1}{2}}{\sqrt{V^2_{int}~g_{tt}~g_{yy}~g_{xx}^{d_0-3}-\tilde{c}^2}}\,du+2\int_{u_k}^{u_0}\frac{c~(g_{uu}/g_{tt})^\frac{1}{2}}{\sqrt{V^2_{int}~g_{tt}~g_{yy}~g_{xx}^{d_0-3}-c^2}}\,du.
\end{align}
First, we plot the subsystem sizes $T_{Re}(u_0)$ and $T_{Im}(u_0)$ corresponding to the surfaces with the real and the imaginary areas by varying the turning point $u_0$ in \cref{T_num}. We observe that the lengths of the subsystems change rapidly when $u_0$ is close to $u_k$ whereas a slower change is observed in the near boundary region ($u_0 \gg u_k$).  To compare the sizes of the subsystems, we plot $\frac{T_{Re}}{T_{Im}}$ as a function of $u_0$ in \cref{d_T_num}, and find the ratio greater than $1$ implying $T_{Im}<T_{Re}$ in the region $u_0\sim u_k$. Note that, this is the precise region where the green surfaces in \cref{fig:sol2all} cross each other. However, the ratio decreases rapidly with $u_0$ moving away from the tip of the geometry where $T_{Im}$ dominates over $T_{Re}$. Finally when $u_0$ is very close to the boundary, $\frac{T_{Re}}{T_{Im}}$ converges to 1 which indicates that $T_{Im}$ and $T_{Re}$ tend to become equal at this limit.

\begin{figure}[ht]
	\centering
 \subfigure[ $|\mathcal{\hat S}^T_{Re}|$ (red-dotted) and $\mathcal{\hat S}^T_{Im}$ (blue-dashed) $vs$ $T$.]{\label{tee_num}\includegraphics[width=75mm]{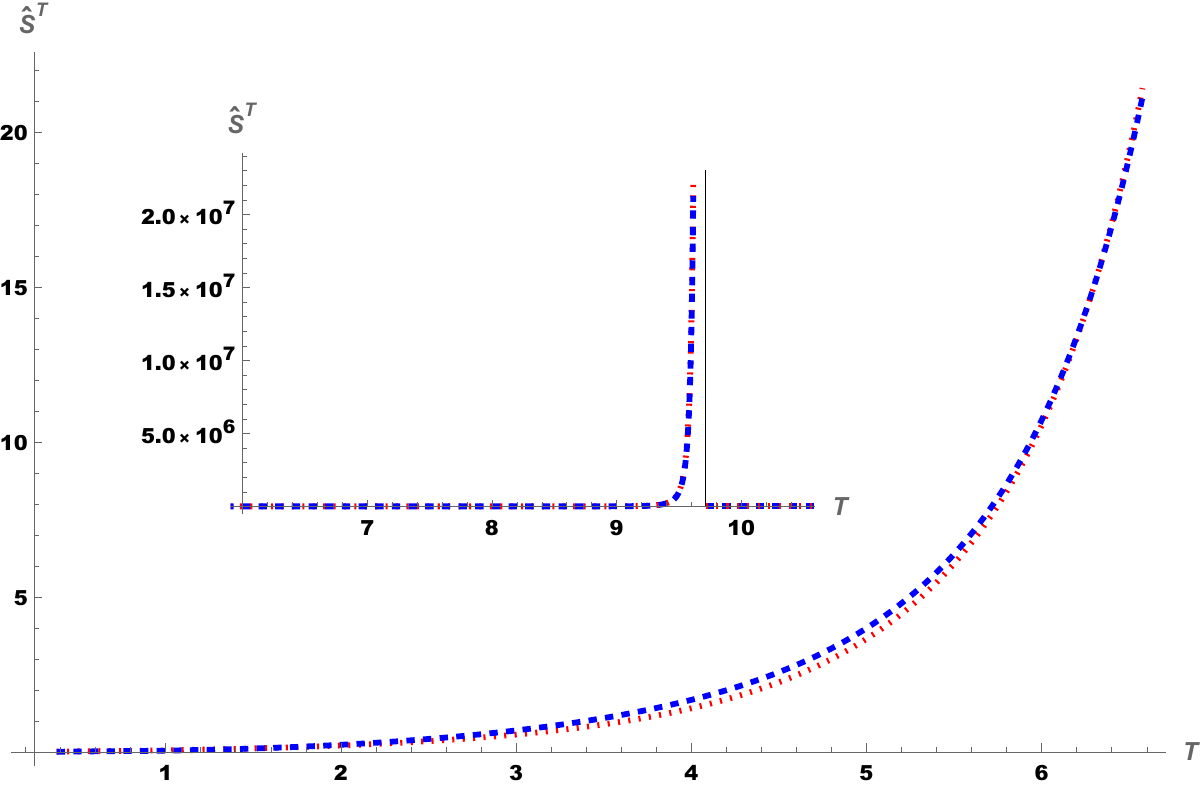}}\hspace{.5cm}
 \subfigure[  $\frac{|\mathcal{\hat S}^T_{Re}|}{\mathcal{\hat S}^T_{Im}}$ $vs$ $u_0$.]{\label{d_tee_num}\includegraphics[width=75mm]{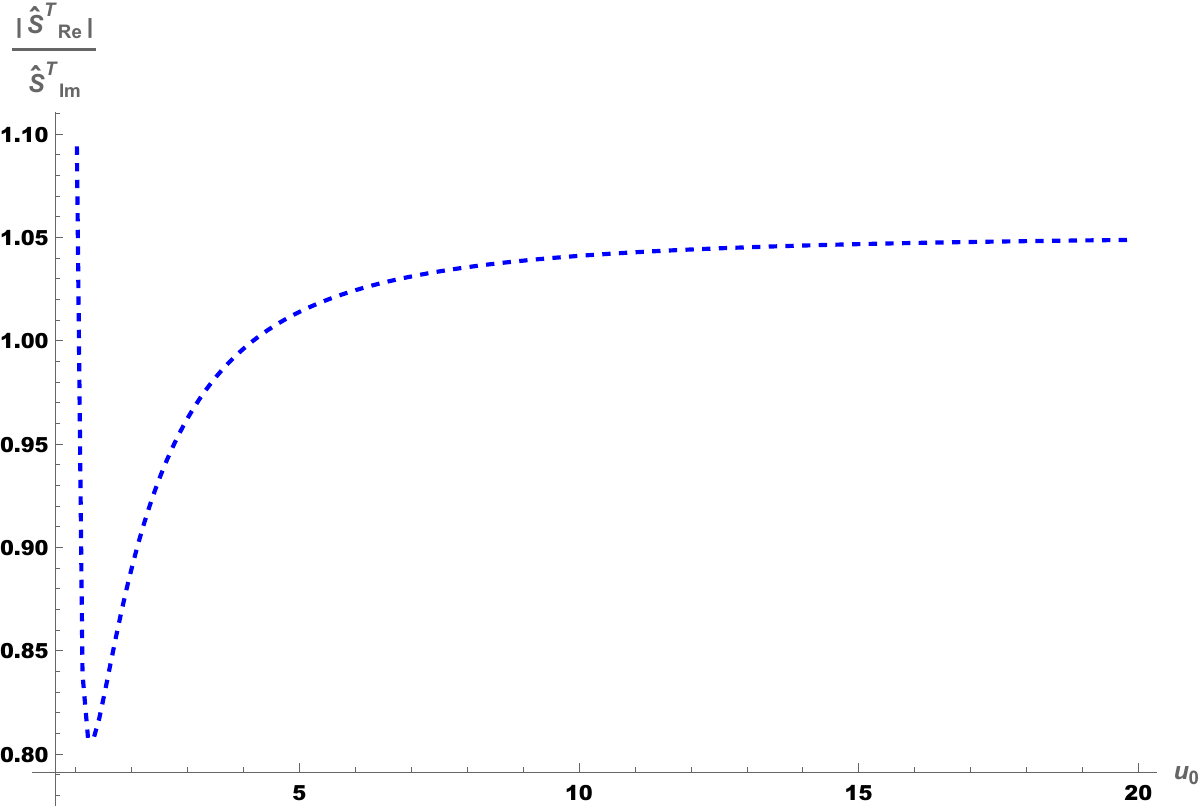} }
	\caption{Analysis of $\mathcal{\hat S}^T$ scaled with $\frac{\cV^{d_0-2}}{4G_N^{d+1}}$ as a function of $u_0$ and $T$. (a) The real and the imaginary parts of $\mathcal{\hat S}^T$ are plotted as a function of $T$ where $T=T_{Re}+T_{Im}$. For small subsystem sizes, both $|\mathcal{\hat S}^T_{Re}|$ and $\mathcal{\hat S}^T_{Im}$ are small whereas they obtain large values where the turning point $u_0$ is far away from $u_k$. In the embedded plot notice the existence of the $T_{crit}$ beyond which a connected non-trivial surface seizes to exist and a discontinuous phase transition occurs. (b)  The ratio of the real and the imaginary parts of tEE in \cref{tgen}. In regions $u_0\approx u_k$ and $u_0\gg u_k$ the real part is larger than the imaginary part. However, there is an intermediate region where the imaginary part dominates. In the near boundary regime   $|\mathcal{\hat S}^T_{Re}|\approx 1.05~\mathcal{\hat S}^T_{Im}$. 
 }
	\label{num2}
\end{figure}

Now, we consider the real and imaginary parts of $\mathcal{\hat{S}}^T$ with respect to the size of the total subsystem $T$ by varying the turning point in \cref{tee_num}. Similar to the results regarding the subsystem sizes, the characteristics of the real and the imaginary parts of the area behave similarly. However, we observe a sign difference between real and imaginary parts. $\mathcal{\hat S}^T_{Re}$ yields a negative value for any set of parameters as $\Scal^T_{discon}>\mathcal{S}^T_{\tilde{c}_0}(u_0)$ in \cref{TEELor2}. We consider $|\mathcal{\hat S}^T_{Re}|$ to obtain a comparison with the $\mathcal{\hat S}^T_{Im}$. In \cref{tee_num}, both $|\mathcal{\hat S}^T_{Re}|$ and $\mathcal{\hat S}^T_{Im}$ increase with increasing size of the total subsystem $T$. It indicates that both the areas increase as the turning point $u_0$ moves towards the boundary. The ratio of the real and the imaginary parts of $\mathcal{\hat S}^T$ is presented \cref{d_tee_num} for comparison. The real part dominates when $u_0$ is in the deep IR or in the near boundary regions, which corresponds to small and large $T$. However, there is an intermediate region where the imaginary part dominates.
Interestingly, in the limit $u_0\gg u_k$, the real and the imaginary parts of $\mathcal{\hat S}^T$ are not exactly equal as the plot in \cref{d_tee_num} shows the convergence of 
$|\mathcal{\hat S}^T_{Re}|/\mathcal{\hat S}^T_{Im}$ approximately at 1.05. This precise value can also be obtained by taking the ratio of the leading terms of \cref{S_renorm_highT2,S_Im_lowT} in the limit $u_0\gg u_k$ which serves as a strong consistency check of the analytic computations presented in the previous section.  The fact that the tEE increases with $T$ can be understood by realizing that $d\mathcal{\hat S}^T_{Re}\sim u_0^{5/2} d T$ for $u_0\gg u_k$,  for the renormalized $\mathcal{\hat S}^T_{Re}$, and that as $u_0$ increases the spacelike real surface moves away from the straight disconnected one, therefore contributing different to the areas.  For $u_0\gg u_k$ the imaginary part $\mathcal{\hat S}^T_{Im}$  increases as well to very large values. This can be understood from the expansions of the previous section and in particular by \cref{S_Im_lowT}. This behavior of $\mathcal{\hat S}^T_{Im}$ is expected since this part of the area is not renormalized and by stretching it towards the boundary its area eventually diverges.

From the numerical analysis we observe that there is a maximum length for $u_0\gg u_k$ that agrees with the analytic expectations and \cref{u0_T_tot}   as $T_{crit}\simeq \g/\sqrt{u_k}$ where the connected surface seizes to exist. For $T< T_{crit}$ we have a non-zero imaginary part $\mathcal{\hat S}^T_{Im}$, as long as the boundary length does not vanish. For larger values of $T>T_{crit}$, there is a phase transition in the observable and the imaginary contribution jumps to zero since then the only solution for this part is the surface $u=u_k$ which makes the union of the surfaces to take the shape of the inverse $\Pi$. Therefore, the existence of the $T_{crit}$ and the behavior described for the tEE makes the pseudoentropy a potential probe of confinement. Moreover, it hints at its capability of signaling other types of phase transitions in quantum theories.

\begin{figure}[ht]
	\centering
        \includegraphics[width=.6\textwidth]{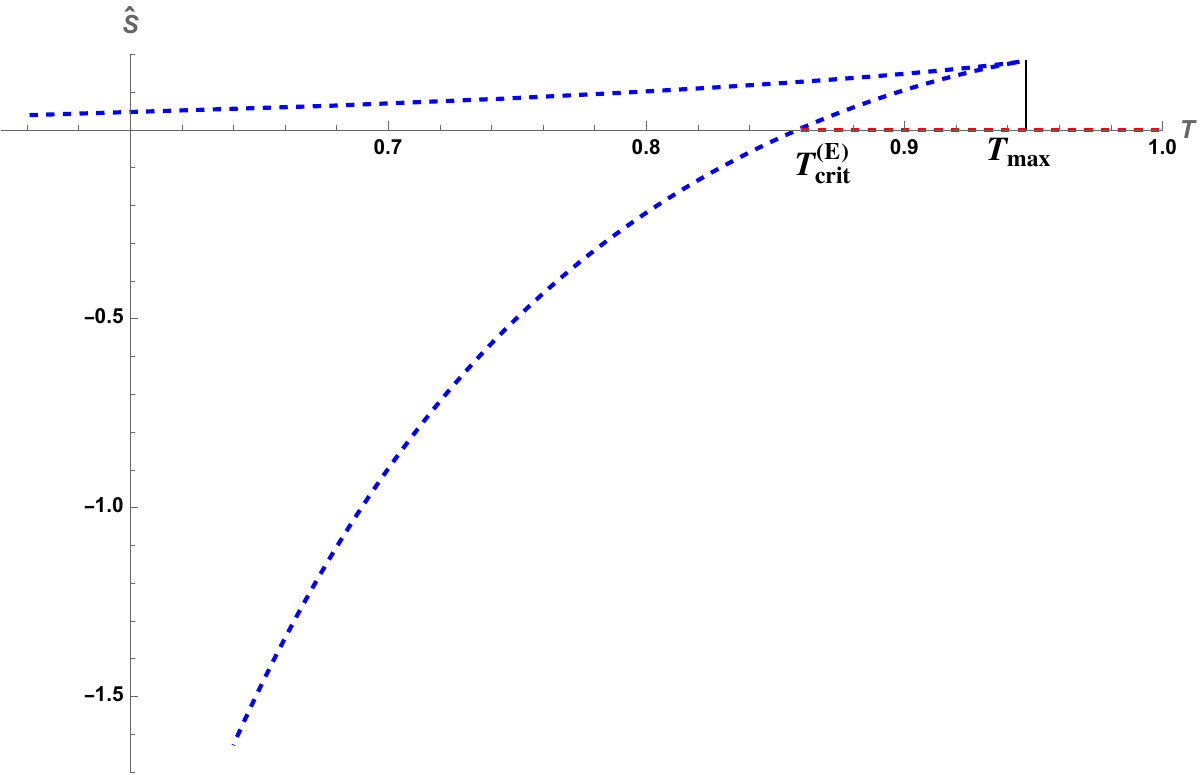}
 \caption{Plot of the EE $\mathcal{\hat S}$ $vs$ $T$ in Euclidean scenario. Here, $\mathcal{\hat S}$ is the difference between the areas of the real part of connected and the disconnected straight surfaces which are two plausible solutions of \cref{gentprime}. The generic expression of $\mathcal{\hat S}$ is given in \cref{genSrenorm}. Note that the plot shows two branches where the lower branch represents the favorable solution.  Following this solution, we observe a change in dominance between the two solutions of \cref{gentprime} at $T=T^{(E)}_{crit}$ where $\mathcal{\hat S}=0$. For regions $T<T^{(E)}_{crit}$ and $T>T^{(E)}_{crit}$ the dominant contributions are received from the connected and the disconnected surfaces respectively. Besides, we observe that the subsystem size obtains a bound which is indicated by $T_{max}$. Beyond this particular value, there is no connected solution of \cref{gentprime}.
 }
		\label{ee_num}
\end{figure}

Before we finish this section let us discuss the entanglement entropy of a subsystem of length $T$ on a constant space slice.  
We numerically compute the UV free EE of a temporal subsystem $T$ as demonstrated in \cref{genSrenorm0}.  We plot the results in \cref{ee_num}, of the EE with respect to the subsystem length $T$. The findings indicate a change in the dominance of the candidate surfaces, where the connected surface dominates for $\mathcal{\hat S}<0$ and the disconnected one dominates for $\mathcal{\hat S}>0$. We consider the surface with minimum area to be the correct analogue of the Ryu-Takayanagi surface. Besides, we also note a bound on the subsystem length beyond which the entanglement entropy does not exist. As a result, we observe the length of the subsystem to be infinitesimally small in both the limits $u_0\to u_k$ and $u_0\gg u_k$ corresponding to the upper and lower branches of \cref{ee_num} respectively. These are well known observations of the EE in confining theories \cite{Klebanov:2007ws}, and the similarity can be understood easily as the metric components $g_{tt}$ and $g_{xx}$ of the confining background are identical in the Euclidean scenario. If we naively analytically continue the length of the EE, we will have a similar phase transition with the original EE, reminiscent of the behavior of the solution $T(u_0)$ in the EE. This simple analytical continuation of the EE will still have a maximum length $T^{(E)}_{crit}$ beyond which a connected solution does not exist, and the solution will consist only of the two straight lines and the part that wraps the tip of the geometry, with a vanishing contribution. Although the quantitative details and specifics differ from the tEE computed directly in Lorentzian signature there are obvious qualitative similarities regarding the phase transition.

\section{Discussions}

In this work, we introduce a holographic definition of the tEE in non-conformal theories, incorporating the influence of a non-trivial dilaton field. We then apply this holographic formula to the confining near-horizon limit of D4-branes with one direction compactified on a circle with anti-periodic boundary conditions for the fermions. Using the Lorentzian signature, we compute the tEE by determining the extremal spacelike and timelike surfaces.  Our results include numerical computations as well as analytical ones. There are two limits where we work with analytical expressions: the limit of a small boundary length which corresponds to the timelike surface staying entirely around the tip of the IR geometry, and a limit where the turning point of the timelike surface reaches near the boundary of the spacetime and corresponds to larger boundary lengths.

We find that in confining systems, the tEE has a non-zero imaginary contribution and exhibits a maximum subsystem length beyond which a connected surface cannot be defined. Beyond this critical length, the connected surface becomes trivial, taking the shape of an inverse $\Pi$, where the imaginary part lies entirely on the tip of the geometry and thus contributes nothing to tEE. This unique behavior of the tEE characterizes confinement, suggesting it can serve as an order parameter for confinement/deconfinement phase transitions. We also show that a simple analytic continuation of the Euclidean EE shares qualitative similarities with the tEE, although the quantitative specifics of the two approaches differ.

Our proposal determines the boundary length solely through the bulk construction of holographic surfaces. In the initial proposal of tEE for conformal theories, the boundary length is fixed prior to the analytic continuation. On the other hand, working exclusively with holographic surfaces, a constant shift in the boundary length is allowed since timelike and spacelike surfaces extend at infinite time distance in the bulk, and the bulk data does not entirely fix the boundary endpoints of the surfaces unless additional information is used. 
In our proposal, we identify a homologous, continuous smooth surface to the boundary subregion $A$ in such a way that fixes the boundary length uniquely. For confining theories, this surface consists of a set of spacelike surfaces that extend all the way from the boundary of the spacetime to the IR holographic wall, which in our spacetime is the tip of the geometry, and a timelike surface that initiates from the IR tip of the geometry, extends towards the boundary direction until its turning point and returns to the IR tip, exhibiting mirror symmetry. The spacelike and timelike surfaces can be naturally merged only at the IR wall at the tip of the geometry and we do that by requiring the extra condition of a well defined first derivative. The process defines a unique homologous surface, as in \cref{fig:sol2}, that specifies uniquely the boundary length, i.e., fully fixes the constants of integration in the equations of motion.

Our paper contributes to advancing the understanding of tEE in non-conformal theories, providing insights into its computation and implications for phase transitions. It would be interesting to extend the tEE computation to black hole environments. The timelike entanglement entropy appears capable of probing the interior of the black hole. It is worthy to examine whether the analytic continuation of the Euclidean EE in the black hole background corresponds to the solitonic tEE, where the boundary subregion is along the compactified direction and if so, how the information of the interior of the black hole translates to the confining case. Moreover, it would be interesting to study the behavior of the tEE, and of the EE in Euclidean time and its analytic continuation, along a renormalization group flow. To what type of degrees does their coarse graining correspond and the existence of monotonicity theorems related to these types of entropies.  In two-dimensional CFT, the imaginary part of the tEE is a characteristic of the theory and depends only on the central charge. The imaginary part in two dimensions can be used to compute the central charge of the theory without any need for numerical fitting with the boundary subregion length; a numerical approach of a tensor network application in spin chain models would be feasible to demonstrate this. Moreover, a c-function as a derivative of the total tEE with respect to the boundary length can be introduced in the same way that was defined for the standard EE, $c:=T ~\frac{\partial \cS^{T}}{\partial T}$. It would be even interesting to isolate the imaginary part of the tEE and examine when and whether a monotonic c-function can be defined explicitly from it. Overall, exploring timelike surfaces and geodesics can open up new avenues for holographic applications in Lorentzian signature. Appropriate extensions of the formalism, as described in our manuscript, allow for smooth surfaces connecting timelike data on the boundary by mixing in the bulk spacelike and timelike segments, enabling connections between boundary and bulk data in novel ways that are useful for a wide variety of related holographic applications.

\section*{Acknowledgment}
The research work of MA and JKB is supported by the National Science and Technology Council of Taiwan with grant 113-2636-M-110-006. The research work of DG is supported by the National Science and Technology Council (NSTC) of Taiwan with the Young Scholar Columbus Fellowship grant 113-2636-M-110-006. 

\appendix

\section{Holographic Time-like Entanglement Entropy in AdS$_3$} \label{section:app1}
One may wonder how our formalism reflects in conformal field theories and the AdS spacetime. In this appendix, we derive tEE for a timelike subsystem in CFT$_2$ at zero temperature which has a bulk dual described by pure AdS$_3$ geometry. We consider the interval, $A= -T/2 < t < T/2$ of length $T$ in the time direction $t$ on $x=0$ slice. The pure AdS$_3$ geometry can be expressed in the Poincar\'e coordinate as,
\be\la{poin}
ds^2=\ff{dz^2-dt^2+dx^2}{z^2}~.
\ee
The minimal area integral can be written as,
\be\la{ads3_area}
\mathcal{S}^T(A)=\ff{1}{4G_N}\int \sqrt{\frac{1}{z^2}-\frac{t'(z)^2}{z^2}} \, dz~.
\ee
The equation of motion from \cref{ads3_area} reads as,
\be\la{ads3_eom}
t'(z)=\pm\frac{c_1 z}{\sqrt{c_1^2 z^2+1}}~,
\ee
where $c_1$ is a constant. Depending on the signature of $c_1^2$, the equation of motion indicates two different types of extremal surfaces. For $c_1^2<0$, the surface corresponding to \cref{ads3_eom} possesses a turning point at $z=z_0$ where $t'(z_0)=\infty$ and $c_1^2=-z_0^{-2}$. We will call these solutions $t_{Im}(z)$ as in the later part of this appendix it will be shown that the area of these surfaces yields imaginary values. Another possibility $c_1^2>0$ does not show any turning point of the surface. We call this solution $t_{Re}(z)$. We demand our conditions described in the main text to hold, which includes a well defined first derivative while "merging" the surfaces. At $z\rightarrow \infty$ we have for the $t_{Re}'(\infty)=\pm 1=t_{Im}'(\infty)$, and we have to connect the branch of the $t_{Re}(z)$ with that of $t_{Im}(z)$ that have the same derivative in order to define the boundary length. Notice that now $g^{\m\n}\pp_\m \S_{Re}\pp_\n \S_{Re}|_{z\rightarrow \infty}=-g^{\m\n}\pp_\m \S_{Im}\pp_\n \S_{Im}|_{z\rightarrow \infty}=-z_0^2$, where $\S$ denote the corresponding hypersurfaces, and surfaces are now at $\infty$, so they are not patched with the strict meaning. According to our prescription, the merging is done for $t_{Re}'(\infty)=t_{Im}'(\infty)$ which gives for the boundary length of the surface
\begin{align}\la{ads3_T}
T_1&=2\int_{z_0}^\infty t_{Im}'(z) dz -2\int_0^\infty t_{Re}'(z) dz 
=2z_0~.
\end{align}
Therefore, for AdS our smoothness requirement imposed on the surface requires a subtraction of the two lengths  $t_{Im}(z)$ and $t_{Re}(z)$.  

Moreover, note that the alternative discarded choice of merging the surfaces as
\begin{align}\la{ads3_T2}
T_2&=2\int_{z_0}^{1/\epsilon_1} t_{Im}'(z) dz +2\int_0^{1/\epsilon_1} t_{Re}'(z) dz =\frac{4}{\epsilon_1}-2z_0~,
\end{align}
where $\epsilon_1\ll1$ introduced to keep track of the infinite result, leads to an infinite length $T_2$ and violates the natural conditions we have imposed on the smoothness of the surface since $t_{Re}'(\infty)\neq t_{Im}'(\infty)$. Furthermore, leads to a problematic $\mathcal{S}^T_{Re}(T)$ that is $T-$independent, despite giving the expected answer for the $\mathcal{S}^T_{Im}$. Therefore, the fact that we have to discard the $T_2$  as the boundary length following our natural conditions for the holographic surfaces is consistent with the expectations for the tEE.

Let us compute the area of these surfaces to obtain tEE. First, we start with the surface where $c_1^2>0$ as,
\begin{align}\label{ads3_STre_fin}
\mathcal{S}^T_{Re}(T_1)&=\ff{1}{2G_N}\int_\epsilon^{\infty} \sqrt{\frac{1}{z^2}-\frac{t'_{Re}(z)^2}{z^2}} \, dz=\frac{1}{2G_N}\log\frac{2 z_0}{\epsilon }=\frac{1}{2G_N}\log\frac{T_1}{\epsilon },
\end{align}
where in the second equality we used $c_1^2=z_0^{-2}$ and \cref{ads3_T} has been imposed to obtain \cref{ads3_STre_fin}. In the above expression, $\epsilon$ is the UV cut-off which appears from the lower limit of the integration indicating an infinite proper length of the geodesic approaching the asymptotic boundary.  
Now we compute the area of the other surface extended from the turning point to the deep IR region. Here we fix the constant $c_1^2=-z_0^{-2}$ and the area integral can be computed as,
\begin{align}
\mathcal{S}^T_{Im}&=\frac{1}{2G_N}\int_{z_0}^\infty \sqrt{\frac{1}{z^2}-\frac{t'_{Im}(z)^2}{z^2}} \, dz=\ff{i\pi}{4G_N}\label{ads3_STim_fin}.
\end{align}
Considering both the contributions, the total tEE can be expressed as a sum of $\mathcal{S}^T_{Re}$ and $\mathcal{S}^T_{Im}$,
\begin{align}
\mathcal{S}^T(T_1)&=\ff{1}{2G_N}\log\frac{T_1}{\epsilon }+\ff{i\pi}{4G_N}\label{ads3_ST}~.
\end{align}
Then it straightforwardly follows using \cite{Brown:1986nw}, that \cref{ads3_ST}  is equal to
\begin{align}
\mathcal{S}^T(T_1)=\ff{c}{3}\log\frac{T_1}{\epsilon }+i\ff{c\pi}{6}~,
\end{align}
which is identical to \cref{res1}, and the results obtained in \cite{Doi:2023zaf,Doi:2022iyj}, justifying the consistency for conformal theories between the methodology described in this manuscript and the prior literature.

\bibliographystyle{JHEP}

\bibliography{timelike}

\end{document}